\documentclass[aps,prd,twocolumn,groupedaddress,nofootinbib]{revtex4-1}
\usepackage[pdftex]{graphicx}
\usepackage{subfigure}
\usepackage{sidecap}

\usepackage{amsmath}
\usepackage{hyperref}
\usepackage{comment}

\usepackage{verbatim}
\usepackage{tikz}
\usepackage{color}

\def\prd{PRD}
\def\apj{ApJ}                 
\def\apjl{ApJL}               
\def\mnras{MNRAS}             

\usepackage{hyperref}
\hypersetup{
    pdfnewwindow=true,      
    colorlinks=true,       
    linkcolor=blue,          
    citecolor=cyan,        
    filecolor=blue,      
    urlcolor=blue           
}

\graphicspath{{pic/}{plot/}}

\def\J{\sc J}
\def\a{a}
\def\omegaz{\Omega}
\def\H{H}
\def\Msun{{M_{\odot}}}

\begin{document}

\title{Black Hole Pulsar}

\author{Janna Levin}
\affiliation{Barnard College of Columbia University, New York, NY 10027, USA }
\affiliation{Chair and Director of Sciences, Pioneer Works, 159 Pioneer St., Brooklyn, NY 11231, USA}
\author{Daniel J. D'Orazio}
\affiliation{Department of Astronomy, Harvard University, 60 Garden Street Cambridge, MA 01238, USA}
\author{Sebastian Garcia-Saenz}
\affiliation{Sorbonne Universit\'es, UPMC Univ.\ Paris 6 and CNRS, UMR 7095, Institut d'Astrophysique de Paris, GReCO, 98bis boulevard Arago, 75014 Paris, France}

\begin{abstract}
In anticipation of a LIGO detection of a black hole/neutron star merger, we expand on the intriguing possibility of an electromagnetic counterpart. Black hole/Neutron star mergers could be disappointingly dark since most black holes will be large enough to swallow a neutron star whole, without tidal disruption and without the subsequent fireworks. Encouragingly, we previously found a promising source of luminosity since the black hole and the highly-magnetized neutron star establish an electronic circuit -- a black hole battery. In this paper, arguing against common lore, we consider the electric charge of the black hole as an overlooked source of electromagnetic radiation. Relying on the well known Wald mechanism by which a spinning black hole immersed in an external magnetic field acquires a stable net charge, we show that a strongly-magnetized neutron star in such a binary system will give rise to a large enough charge in the black hole to allow for potentially observable effects. Although the maximum charge is stable, we show there is a continuous flux of charges contributing to the luminosity. Most interestingly, the spinning charged black hole then creates its own magnetic dipole to power a black hole pulsar.
\end{abstract}


\maketitle

\bigskip

\section{Introduction}

The LIGO collaboration recently announced the first detection of gravitational
waves from a neutron star (NS) collision \cite{TheLIGOScientific:2017qsa}.
Stepping on the heels of the gravitational wave train, all manner of fireworks
are anticipated when the dense neutron-star matter crushes together.
Anticipations were beautifully confirmed since the FERMI and INTEGRAL
satellites detected a gamma-ray burst from the same direction
\cite{Monitor:2017mdv,Goldstein:2017mmi,Savchenko:2017ffs}. Over the next two
weeks, dozens of instruments and a significant fraction of the astronomical
community directed their focus and witnessed pyrotechnics in the aftermath
across the electromagnetic (EM) spectrum \cite{GBM:2017lvd}. The era of
multi-messenger astronomy has begun spectacularly.

At the other extreme, black hole (BH) collisions are expected to be
spectacularly dark. The LIGO BH mergers exhibited no detectable
electromagnetic counterpart, although there were intriguing gamma-ray
signatures from near GW150914 and GW170104 (\citep[][and see
\citet{DOrazioLoeb:2018} and references therein]{Connaughton+2016,
AGILE:Verrecchia+2017, Connaughton+2018_Fermi2}, that may or may not have been
correlated with the gravitational-wave events.  BHs are empty space and their
merger will be invisible, unless dressed in ambient debris. The BH collisions
were the most powerful events detected since the big bang and yet it is
possible that none of the energy came out in the electromagnetic spectrum.
All of the energy emanated in the darkness of gravitational waves.

Next in the compact object combinatorics will inevitably be black hole/neutron
star (BH/NS) collisions. While the tidal disruption of the NS in these systems
could occur for the smallest BH partners, resulting in a short gamma-ray burst
\citep[\textit{e.g.},][]{NPP:1992,ShibataBHNSLRR:2011}, BHs larger than $\sim
8 \Msun$, will swallow the NS whole -- an expectation further endorsed by the
large BHs LIGO observed \citep{LIGO_BBHO1:2016, GW170104:2017, GW170608,
GW170814}. Without tidal disruption, there is not an obvious source of light.

Fortunately, there is another mechanism for the system to light up: the Black
Hole Battery \cite{GLB:1969, McL:2011, DL:2013}. NSs are tremendous
magnets. As they whip around a BH companion, the orbiting magnet
creates a source of electricity. How this electricity is channeled into a
light element remains somewhat uncertain although we have suggested several
viable channels, including synchro-curvature radiation, a fireball,
and a fast radio burst \cite{DL:2016, MingLevin+2015}.

In this article we argue that another largely overlooked EM channel requires
further exploration: BH charge. Historically, a dismissive argument
has been made that a charged BH will discharge essentially
instantaneously, the electromagnetic force being so excessively strong. Any
errant charges will easily and swiftly be absorbed from the interstellar
medium to counter the charge of the BH, the argument goes. However, as
shown in an elegant paper by Wald in 1974 \cite{Wald:1974np}, a BH
immersed in a magnetic field actually favors charge energetically. In other
words, the BH simply will acquire stable charge if it spins in a
magnetic field. We therefore expect a BH battery -- a BH
pierced by the field lines of an orbiting NS magnet -- to acquire a
significant charge of the Wald value, $Q_W=2 B_o a M$  where $B_o$ is the
strength of the NS dipole field at the location of the BH of
mass $M$ and $a$ is the spin of the hole. Since magnetic dipoles drop off
quickly, by $r^{-3}$, the Wald charge is small until the final stages of
merger. See refs.\ \cite{Zhang:2016rli, Fraschetti:2016,Liu:2016olx} for
interesting recent studies of electromagnetic counterparts in charged BH/BH mergers, ref.\ \cite{Nathanail:2017wly} for work on gravitational collapse to a charged BH, and ref.\ \cite{Zajacek:2018ycb} which considered the Wald mechanism applied to the central galactic BH.

The no-hair theorem is often misinterpreted as enforcing zero magnetic fields
on a BH in vacuum. Actually, and more sensibly, the no-hair theorem
ensures that the only magnetic field a BH can support is consistent
with a monopole of electric charge. A spinning electric charge naturally
creates a magnetic dipole. So a spinning charged BH has all of the
attributes of a pulsar: spin, a magnetic field, and a strong electric field to
create a magnetosphere. We predict the formation of a short-lived and erratic
BH pulsar prior to merger that could well survive briefly post-merger
before the magnetosphere and charge dissipate.

The characteristics of the BH pulsar follow from the NS magnetic field. The Wald charge on a BH immersed in an external NS dipole field, which drops off as the cubed distance between the two, $r^{-3}$, would be
\begin{equation}
\begin{aligned}
Q_W& \approx 10^{-7} M
\left (\frac{a}{M}\right )\left (\frac{M}{10 M_\odot}\right )^2 \left (\frac{B_{NS}}{10^{12} G}\right ) \left (\frac{R_{NS}}{r}\right )^3.
\label{Eq:QW}
\end{aligned}
\end{equation}
Here $B_{NS}$ is the NS's magnetic field at the surface of the NS and $R_{NS}$
is the radius of the NS. Note that $r\geq R_{NS}$. At it's maximum, $Q_{W,{\rm
max}}\sim 10^{-7} M$ (which comes to $\approx 10^{24}\,{\rm statCoulombs}$),
so we can still use the Kerr solution. Assuming a NS with a mass of $1.4
\Msun$ and angular spin frequency of $\Omega_{\rm NS}=0.1$ seconds, we find
that when the BH enters the light cylinder of the NS, $r=R_{LC} = c /
\Omega_{\rm NS}$, the charge is 10 billion times smaller, $Q_{LC}\sim 10^{-10}
Q_{W,{\rm max}}$. Over the next $t_{\rm merger}-t_{LC}\sim 3$ years the charge
increases. In the final minute of inspiral, when the binary is emitting at
$\gtrsim 17$~Hz, in the LIGO band, the charge increases by a factor of a
million. As $r\rightarrow R_{NS}$, $Q\rightarrow 10^{33} e$ which is only
about $10^3$ kgs of electrons.

For reassurance that the black hole actually has time to acquire charge, we estimate the charging timescale. 
While there are lots of 
uncertainties in such an assessment, we consider 
an initially vacuum configuration that siphons charge 
from the magnetosphere of the NS.
Then the charging timescale can be estimated as the light crossing time of
the BH/NS system, $r/c$. The ratio of GW inspiral time to the charging time is
$t_{\rm GW}/(r/c) \approx 1.5  (r/(2GM/c^2))^3$ for the fiducial binary values
chosen here, which confirms that the charging timescale is much shorter than the inspiral timescale until
merger. Longer charging timescales could arise in non-vacuum, force-free
magnetospheres \citep[\textit{e.g.},][]{LyutikovMckinney:2011}. However,
because of the $r^3$ dependence in the timescale ratio above, one would need
the charging timescale to be $\mathcal{O}(10^3)$ times longer than the light
crossing time to mitigate the BH charge in the last second of inspiral. This estimate is encouraging, suggesting
that the black hole would have time to charge before merger.

Once charged, 
the spinning black hole supports a magnetic dipole field.
Take the magnetic dipole moment of the BH to be of order 
${\sc m} \sim Q_W M$. The BH $B$-field is comparable to, though of course less than, the field in which it's submerged. We can estimate
the magnitude of the $B$-field as $B_{BH} \sim m/r^3$. Then using $Q_W=2B_oaM$ with $B_o$ given by the dipole field of the NS at the location of the BH,
\begin{equation}
B_{BH} = \frac{1}{2}  \left (\frac{a}{M}\right )^2 B_{NS} \left( \frac{R_{NS}}{r}\right)^{3} \left( \frac{2M}{r}\right)^{3}
\end{equation}
One factor of $a$ determines the
magnitude of the Wald charge while the other determines the magnitude of the
magnetic moment sourced by the spinning, charged BH. Pulsars are hard to see
far away (\textit{i.e.}\ outside of the galaxy), so we consider other channels
for luminosity than just the BH pulsar.

In addition to the BH pulsar, we suggest that the flux of charge around the BH will create significant luminosities potentially detectable for the range of instruments in the LIGO network. There are two clear opportunities for particle acceleration: At the moment the BH charges up pre-merger and the moment the BH discharges post-merger. A third interesting possibility is the continual fluxing of charges within the magnetosphere. Although the Wald charge appears to be stable, negative and positive charges continue to course along field lines since in vacuum $E\cdot B\ne 0$. And, as we discuss in \S \ref{S:ParticleAcceleration}, there is no value of the charge for which $E\cdot B=0$ everywhere.

As an order of magnitude estimate, we calculate the total power that could be released
if a fraction $f$ of the power associated with the Wald charge in the
Wald electric field, $E_W$, were released,
\begin{equation} 
\nonumber
f Q_{W}E_{W} c \approx 2 \times 10^{45} {\rm{erg} \ \rm{s}^{-1} }
f \left(\frac{B_{NS}}{10^{12} \rm{G}}\right)^2 \left(\frac{R_{NS}}{r} \right)^6 \left(\frac{M}{10 \Msun}\right)^2.
\end{equation} 
where, for the electric field, we use the horizon Wald
electric field at the poles, within an immersing magnetic field corresponding to a NS with surface mangetic feld $B_{NS}$, at a distance of $3R_{NS}$, 
\begin{equation}
E_W \approx 3.3
\times 10^{10} \rm{statV/cm} \left( \frac{B_{NS}}{10^{12}\rm{G}}\right ) \\ .
\end{equation}
This is of order the largest electric field achievable in the system and will
decrease for larger BHs that cannot approach as closely the magnetic field of
the NS.

Now, it's fair to expect that given the large electric fields involved, the
BH will create its own magnetosphere
\citep[\textit{e.g.},][]{RudSuth:1975, Blandford:1977ds}, as well as enter the
magnetosphere of the NS. As the system transitions from vacuum to
force-free, the Wald argument no longer holds. Do force-free BH
systems also have charge and regions of particle acceleration, as a neutron
star pulsar does? That remains an open question that we intend to investigate
in full numerical general relativity. Compellingly, we do show that even
the classic Blandford-Znajek solution has a small charge. It's also worth noting that the Goldreich-Julian pulsar \cite{Goldreich:1969sb} is force-free and charged \cite{1975ApJ...201..783C}.

Before we proceed, a quick comment on notation. Where unambiguous, we'll suppress index notation and use a $\cdot$ to indicate a sum over $4$-indices. Between vectors this is unambiguous. For tensors, the order determines the index to be summed. By example, for 2-tensors (or pseudo-tensors) $H$ and $K$, $H\cdot K$ sums the final index of $H$ over the first index of $K$. Explicitly $H\cdot K=H_{\alpha \mu}K^{\mu \beta}$. The placement of the free indices up or down is ambiguous in this notation. A double $\cdot \cdot$ means $H\cdot \cdot K=H_{\alpha \mu} \cdot K^{\mu \beta}=H_{\alpha \mu} K^{\mu \alpha}$. We'll resort to explicit indices as required in context.  

We'll work as generally as possible but when the time comes to restrict to the particular Kerr metric, we use Boyer-Linquist coordinates:
\begin{eqnarray}
ds^2 = &-& \left (1-\frac{2Mr}{\Sigma}\right ) dt^2 +\frac{\Sigma}{\Delta} dr^2 + \Sigma d\theta^2 \nonumber \\
&+& \frac{(r^2+a^2)^2-\Delta a^2\sin^2\theta}{\Sigma}\sin^2\theta d\phi^2 \nonumber \\
&-& \frac{4Mar\sin^2\theta}{\Sigma} dt d\phi\,,
\label{Eq:BL}
\end{eqnarray}
with
\begin{eqnarray}
\Sigma & =& r^2 + a^2\cos^2\theta\,, \nonumber \\
\Delta &=& r^2+a^2 - 2Mr\,.
\end{eqnarray}
There are a number of useful metric quantities that greatly ease calculations and that we compile in Appendix \ref{App:A}.

The paper is outlined as follows. In Section \ref{S:Review of Wald's
Argument}, we review Wald's argument for the charging up of a Kerr BH in a
uniform magnetic field (the Wald solution). In Section \ref{S:Equations of
Motion} we present the equations of motion for test charges in the Wald
solution. In Section \ref{S:Charge
Accretion} we consider charge accretion in the Wald solution, at the poles of
the BH, and its need for generalization to charge accretion in the global
spacetime. Section \ref{S:ParticleAcceleration} presents numerical solutions
to the equations of motion for test charges in the Wald fields addressing
the question of global charge accretion. Section \ref{S:ParticleAcceleration}
also considers EM emission from the acceleration of test charges in the Wald
field. Section \ref{S:FF} briefly considers BH charge in the force
free limit. Section \ref{S:Summary} concludes.

\section{Review of Wald's Argument}
\label{S:Review of Wald's Argument}

We begin with Wald's elegant EM solution around a spinning BH immersed in a magnetic field that is uniform at infinity \cite{Wald:1974np}. The generalization including the backreaction of the EM field on the geometry has been studied in \cite{Dokuchaev:1987ova,Gibbons:2013yq}; see also \cite{Morozova:2013ina} for a related analysis of a moving BH. The vacuum Maxwell equations are
\begin{equation}
D\cdot F=0
\end{equation}
for $F=dA$, where $d$ is the usual exterior derivative and $A$ is the vector potential.
Imposing the Lorentz gauge $D\cdot A =0$, Maxwell's equations for the vector potential become
\begin{equation}
\left (D\cdot D \right )A = 0 \,. \label{Eq:ME}
\end{equation}
Wald's solution leverages the Killing vectors $\psi$ and $\eta$ that correspond to the axial symmetry and the stationarity of the Kerr spacetime respectively. 
Killing vectors satisfy Killing's equation
$D_{(\mu}\psi_{\nu)}=0 $, which we massage into a new form after 
taking another covariant derivative
\begin{equation}
D^\mu D_{(\mu}\psi_{\nu)}=
D^\mu D_\mu \psi_\nu +D^\mu D_\nu \psi_\mu =0\,.  \label{Eq:DKE}
\end{equation}
We swap the order of the derivatives in the second term on the LHS using
\begin{equation}
D_\mu D_\nu  \psi^\mu=D_\nu D_\mu  \psi^\mu +R_{\mu \nu}\psi^\mu \,.
\end{equation}
For the vacuum Kerr solution $R_{\mu \nu}=0$ and Killing's equation ensures
$D \cdot \psi = \frac{1}{2}g^{\mu\nu}D_{(\mu}\psi_{\nu)}=0$, which together render
$D_\mu D_\nu \psi^\mu =0$.
Consequently, Eq.\ (\ref{Eq:DKE}) is just
\begin{equation}
\left (D\cdot D \right )\psi = 0\,,
\end{equation}
which is precisely Eq.\ (\ref{Eq:ME}), Maxwell's equations for the EM vector potential in Lorentz gauge.
Beautifully, the Killing vectors are automatically solutions of Maxwell's equations. 
The vector potential is then a linear sum of the Killing vectors $\psi$ and $\eta$ with constant coefficients. 

Using Gauss's Law and the geometric interpretation of the Killing vectors \cite{Wald:1974np}, the coefficients can be chosen to find the uncharged solution that asymptotes to a uniform magnetic field $B_o$,
\begin{equation}
A=\frac{1}{2}B_o\left (\psi + 2a \eta   \right ) \,.
\label{Eq:WaldA}
\end{equation}
And in a few short steps we have the full EM solution for an uncharged BH of spin $a$ aligned with an otherwise uniform magnetic field $B_o$.

\section{Equations of Motion}
\label{S:Equations of Motion}

To track the motion of charged particles we begin with the super-Hamiltonian in terms of the canonical momentum $\pi$ \cite{Misner:1974qy}
\begin{equation}
\H=\frac{1}{2}\left ( \pi -q A \right ) \cdot \left ( \pi -q A \right ) \,.
\end{equation}
The 4-velocity  is defined as $u=\dot x$, where a dot denotes differentiation with respect to proper time $\tau$ so that $u\cdot u =-1$.
We also define $p=mu$ so that $p\cdot p= - m^2$.
The first of Hamilton's equations gives 
\begin{equation}
p = \frac{ \partial {\H}}{\partial\pi} =  \left ( \pi -q A \right ) \,.
\end{equation}
The other of Hamilton's equations yields the equations of motion
\begin{equation}
\left ( p \cdot D\right ) p =  q F\cdot p\,, \label{Eq:Geod}
\end{equation}
where the RHS is the relativistic Lorentz force.

For a stationary, axisymmetric spacetime with a stationary, axisymmetric electromagnetic field, there are two immediate constants of the motion. More formally, for any Killing vector $\psi$, 
if the Lie derivative of the electromagnetic field vanishes,
\begin{equation}
{\mathcal L}_\psi A = \psi \cdot D A - A\cdot D \psi =0\,,
\end{equation}
then the quantity $\pi \cdot \psi$ is conserved along the worldline:
\begin{equation}
p\cdot D(\pi \cdot \psi )=\frac{d}{d\tau}(\pi \cdot \psi )=0 \, .
\end{equation}
Our two Killing vectors yield a conserved energy $\varepsilon$ and a conserved angular momentum $\ell$:
\begin{eqnarray}
\varepsilon &=& -\pi \cdot \eta\,, \nonumber\\
\ell &=& \psi\cdot \pi\,. \label{Eq:EL}
\end{eqnarray}

As Carter usefully showed, for a Killing tensor $K$ there is an associated conserved quantity in the absence of an electromagnetic field, the Carter constant $K\cdot \cdot u u$. Naively we would expect that when $A\ne 0$, that $K\cdot \cdot \pi \pi$ is conserved, if the field respects some suitable restrictions. It is not clear what these restrictions are, as there is no obvious analogue of Lie derivative with respect to a rank-2 tensor. In fact, using a method developed by Van Holten \cite{vanHolten:2006xq}, Ref.\ \cite{Igata:2010ny} established that there is no conserved quantity associated with the Killing tensor of the Kerr spacetime whenever the external magnetic field is nonzero (see also \cite{Kolar:2015cha} for a more recent and general analysis). This is further supported by numerical studies of charged particle motion around a magnetized Kerr BH, which evidence chaotic behavior and hence the non-integrability of the equations of motion \cite{Takahashi:2008zh,2010ApJ...722.1240K}.

\subsection{Carter constant}

Although a proof of the absence of a Carter constant exists in the references cited above, we present a simple little argument here that suggests another route to the proof.

Carter showed that for a Hamiltonian of the form
\begin{equation}
H=\frac{H_r+H_\theta}{2(U_r+U_\theta)} \, ,
\label{Eq:Hform}
\end{equation}
where $U_r$ is solely function of $r$, $U_\theta$ is solely a function of $\theta$, $H_r$ is a function of $r$ and all $\pi$'s except $\pi_\theta$, and $H_\theta $ is a function of $\theta$ and all $\pi$'s except $\pi_r$,
there exists a 
\begin{equation}
K = \frac{U_r H_\theta - U_\theta H_r}{(U_r+U_\theta)}
\end{equation}
such that the Poisson bracket vanishes:
\begin{equation}
 \{K,H \}=\frac{\partial K}{\partial x^i}\frac{\partial H}{\partial \pi_i}-\frac{\partial K}{\partial \pi_i}\frac{\partial H}{\partial x^i}=0 \, .
 \label{Eq:K}
\end{equation}
In other words, $K$ is a constant of motion. 
The proof goes like this. We rewrite
\begin{equation}
K=2U_rH-H_r
\end{equation}
Then
\begin{equation}
\{H,K \}=2\{H,U_r\}H-\{H,H_r\} \, .
\end{equation}
By design
\begin{equation}
\begin{split}
\{H_r,H_\theta \} & =0 \,, \\
\{U_r,U_\theta \} & =0 \, .
\end{split}
\end{equation}
We also note that
\begin{equation}
\{H,H_r \} =-\frac{H}{(U_r+U_\theta)}\{U_r,H_r\}=-2H\{U_r,H\} \, .
\end{equation}
Using these in the original Poisson bracket,
we quickly get that
\begin{equation}
\{H,K\}= 0 \, ,
\end{equation}
and $K$ is conserved. We could equally well have written $K=-2U_\theta H +H_\theta$ and followed through to the same conclusion.

For a charged, Kerr BH, the vector potential is just
\begin{equation}
A= -\frac{Q}{2M} \eta
\end{equation}
and the Hamiltonian becomes
\begin{equation}
H=\frac{1}{2}(\pi - qA)\cdot (\pi - qA)=\frac{1}{2}(\pi\cdot \pi -2q \pi \cdot A +q^2 A\cdot A) \, ,
\end{equation}
which this has the form of Eq.\ (\ref{Eq:Hform})
with 
\begin{equation}
\begin{split}
H_r =
\Delta \pi_r^2-&\frac{(r^2+a^2)^2}{\Delta}\pi_t^2 -\frac{a^2}{\Delta} \pi_\phi^2+
\\
-&\frac{4Mar}{\Delta}\pi_t\pi_\phi
+\frac{qQ}{M}r^2\pi_t
-\frac{q^2Q^2}{4M^2}\Delta
\\
H_\theta  =
\pi_\theta^2+&a^2\sin^2\theta \pi_t^2+\frac{1}{\sin^2\theta}\pi_\phi^2
\\
+&\frac{qQ}{M}a^2\cos^2\theta \pi_t +
\frac{q^2Q^2}{4M^2}a^2\sin^2\theta
\\
U_r +
U_\theta &=\Sigma.
\end{split}
\end{equation}
So the charged, Kerr solution has a conserved $K$.

However, when the vector potential has the form
\begin{equation}
A=c_t \eta + c_\phi \psi,
\end{equation}
as it does in our setting, then the Hamiltonian has the form
\begin{equation}
H=\frac{H_r+H_\theta + H_\times}{2(U_r+U_\theta)} \, ,
\label{Eq:Hbad}
\end{equation}
where we replace
\begin{equation}
\begin{split}
-\frac{Q}{2M}& \rightarrow c_t \\
H_r &\rightarrow H_r -2q r^2 c_\phi \pi_\phi \\
H_\theta &\rightarrow H_\theta -2q a^2 \cos^2\theta c_\phi \pi_\phi,
\end{split}
\end{equation}
and
\begin{eqnarray}
H_\times & = & q^2 \Sigma \left ( c_\phi^2 \psi\cdot \psi +2c_\phi c_t \eta \cdot \psi  \right ) \\
&= & q^2 \sin^2\theta\left (c_\phi^2 \left ( (r^2+a^2)^2-\Delta a^2 \sin^2\theta\right ) -c_\phi c_t  4Mar\right )\nonumber ,
\end{eqnarray}
and $H_\times$ is a function of $(r,\theta)$ that is no longer separable.
Suppose we try to find a new constant, $\bar K$, by examining the non-vanishing piece of $\{H,K\}$ for $K=2U_r H -H_r$. If we can rewrite $\{H,K\}=\{H,Z\}$ then we can subtract $Z$ to find a new constant, $\bar K = K -Z$. The non-vanishing piece comes explicitly from the term $\{H,H_r\}$,
\begin{equation}
\{H,K\}= -\frac{1}{2(U_r+U_\theta)}\{H_\times,H_r\} \, .
\end{equation}
We can in fact manipulate this into the form $\{H,Z\}$:
\begin{equation}
\begin{split}
\{H,K\}&=-\frac{1}{2(U_r+U_\theta)}\{H_\times,H_r\} \\
&=\{\frac{H_r}{2(U_r+U_\theta)},H_\times\} \\
&=\{H-\frac{H_\theta}{2(U_r+U_\theta)},H_\times\} \\
&=\{H,H_\times\} +\frac{1}{2(U_r+U_\theta)}\{H_\times,H_\theta\} \\
&=\{H,H_\times\} +\{H,H_\theta\} +\frac{H}{2(U_r+U_\theta)^2}\{U_\theta,H_\theta\} \\
&=\{H,H_\times\} +\{H,H_\theta\} -2H\{H,U_\theta\} \\
&=\{H,H_\times+H_\theta -2U_\theta H \}=\{H,Z\}
\end{split}
\end{equation}
Notice that the Poisson bracket with $Z$ is not zero unless $H_\times =0$.
Subtracting $Z$ from $K$ gives our new constant
\begin{equation}
\bar K = K - Z = 2(U_r+U_\theta) H - (H_r+H_\theta+H_\times) 
\end{equation}
but this is identically zero.
In other words, we have lost our Carter constant and the equations are anticipated to be non-integrable, permitting chaotic behavior.

Granted, the above argument lacks the compelling feature of the uniqueness of $\bar K$, which we haven't proven. 
And this might even seem like a slight of hand. But notice that this method would have led to the correct form for $K$ in the charged Kerr solution. Start with $-H_r$. Take $\{H,-H_r\}$ with $H_\times=0$ and re-express as $\{H,Z\}$:
\begin{equation}
\begin{split}
\{H,-H_r\}&=\frac{H}{(U_r+U_\theta)}\{U_r,H_r\} \\
&=-2H\{\frac{H_r}{2(U_r+U_\theta)},U_r\} \\
&=-2H\{H,U_r\} \\
&=\{H, -2U_r H \}=\{H,Z\}
\end{split}
\end{equation}
to find $K=-H_r-Z=2U_rH-H_r$, which is Eq.\ (\ref{Eq:K}) as promised. 

Notice, we do have a Carter constant in the equatorial plane because $H_\times$ becomes separable when $\theta$ is constant at $\pi/2$ 
and for radial motion along the poles because $H_\times=0$ when $\theta$ is constant at $0$ and $\pi$. 

\section{Charge Accretion}
\label{S:Charge Accretion}

Now here is where the situation gets interesting for us in the astrophysical context. The uncharged solution is unstable to the acquisition of charge. Wald demonstrates that a positive charge released from infinity along the pole will be accreted onto the BH and a negative charge will be repelled (if the BH spin aligns with $B_o$, and the reverse if the spin is anti-aligned). Without loss of generality, we assume aligned spin in the discussions.

Wald's argument, based on energetics, as interesting as it is, restricts to the poles and is not transparently covariant. Carter showed that the electrostatic potential for a ZAMO (zero angular momentum observer) is constant on the horizon, which means Wald's argument applies off the poles if you ask a ZAMO \cite{Carter:1973rla}. However, as we show that does not equate to $E\cdot B=0$.
We'll run through the dynamics on the symmetry axis before delving into the implications of generalizing off the poles.

Lower a charged test particle with charge $q$ down the axis of symmetry along the poles ($\theta=0,\pi$) from infinitely far away to the horizon at $r=r_{+}$. The conserved energy is
\begin{equation}
\varepsilon = -\pi \cdot \eta = -(p\cdot \eta + q A\cdot \eta) \, .
\end{equation}
The first term is the kinetic energy, according to an observer on the worldline $u=\eta$, and the second term is an electrostatic potential energy. Focusing on the second term,
the change in the electrostatic energy of the particle at the horizon versus at infinity is
\begin{equation}
\delta \varepsilon =\left. -qA\cdot \eta\right |_{r_+}+\left. qA\cdot \eta\right |_{\infty} \, .
\end{equation}
From this and the electromagnetic four-potential for the fields around an uncharged BH (Eq. \ref{Eq:WaldA}) we find a so-called injection energy 
\begin{eqnarray}
\delta \varepsilon &=& \left.- qA\cdot \eta\right |_{r_+}+\left. qA\cdot \eta\right |_{\infty} \nonumber \\
&=&-q\left. \left [ \frac{1}{2}B_o\left (\psi\cdot \eta +2\a\eta\cdot \eta\right ) 
\right ]\right |^{r_+}_{\infty} \nonumber \\
&=& -q\left. \left [ \frac{1}{2}B_o\left (g_{t\phi} +2\a g_{tt}\right )
\right ]\right |^{r_+}_{\infty} \, .
\end{eqnarray}
On the poles $g_{t\phi}=0$, and on the horizon $g_{tt}=0$ while $g_{tt}=-1$ at infinity giving
\begin{equation}
\delta \varepsilon=- qB_o\a\,.
\end{equation}
Since $\delta \varepsilon/q < 0$ for a positive charge, the potential is higher at infinity and lower at the horizon. Intuitively, the electric field, and therefore the electric force on a positive charge, will point from high to low potential. So we therefore expect the BH to accrete positive charges until $\delta \varepsilon/q = 0$. 

A BH of charge $Q$ in a uniform field $B_o$ has electromagnetic four potential
\begin{eqnarray}
A  &=& \frac{1}{2}B_o\psi +\frac{1}{2M}\left  (2B_o\a M -Q\right )\eta\,.
\end{eqnarray}
Running through the same argument for a test charge $q$ lowered from infinity to the event horizon of a charged BH, the change in the electrostatic energy is
\begin{equation}
\delta \varepsilon=q\left [\frac{Q}{2M}- B_o\a\right ]\,.
\end{equation}
For $Q=Q_W\equiv 2B_o\a M$, the energy difference vanishes and the BH has charged up to a stable value.

However, this argument is not explicitly covariant. Only an observer on the worldline $u=\eta$ measures the electrostatic potential as
\begin{equation}
V=-A\cdot \eta\, .
\end{equation}
The set of such (non-inertial) observers cannot fire rockets hard enough when too near the event horizon. In other words, close enough to the BH, there is no such timelike worldline. A stronger argument, which we'll pursue in a subsequent section would be to look for force-free solutions, which require the covariant condition
\begin{equation}
\frac{1}{4}Tr( F\cdot \tilde F)= E\cdot B=0 \, .
\end{equation}
And, in fact, $E\cdot B=0$ along the poles only when the BH has acquired the Wald charge. This confirms the argument that when the BH is charged to $Q_W=2B_o\a M$ particles will no longer experience EM forces along the poles. 

However, this argument does not generalize off the poles. Away from the poles, $A\cdot \eta\ne 0$ since both $\psi \cdot \eta$ and $\eta\cdot \eta$ are non-zero and theta dependent. Consequently, there's no value of $Q$ which kills $\delta \varepsilon$.

Off the poles, we could ask a ZAMO what she sees in terms of the electrostatic energy. A particle has zero angular momentum when $\ell=0$. If we take the particle off a geodesic, set $\dot \theta=0$ and fire rockets so that $ \dot r=0$, then 
\begin{equation}
u_Z=u^t(\eta + \omegaz  \psi)\,,
\end{equation}
and for $q=0$,
\begin{equation}
\ell / m=0 = u_Z\cdot \psi = u_{Z \phi}=u^t  (\psi \cdot \eta +\omegaz  \psi \cdot \psi)\,,
\end{equation}
which fixes 
$\omegaz $ to the ZAMO's angular velocity:
\begin{equation}
\omegaz =-\frac{\psi \cdot \eta}{ \psi \cdot \psi}=-\frac{g_{t\phi}}{g_{\phi\phi}}=\frac{2Mar}{(r^2+a^2)^2-\Delta a^2\sin^2\theta}\, .
\end{equation}
Then $u\cdot u=-1$ fixes $u^t$,
\begin{equation}
(u^t)^2\left ( \eta^2+\omegaz ^2 \psi^2 +2\omegaz  \eta\cdot \psi \right )=(u^t)^2\left ( \eta^2+\omegaz  \eta\cdot \psi \right ) = -1\,.\nonumber
\end{equation}
Expressing this in terms of metric components, and using the useful relations in Appendix \ref{App:A}, we have
\begin{equation}
u^t= \frac{g_{\phi\phi}^{1/2}}{\left (- g_{tt}g_{\phi\phi}+g_{t\phi}  g_{t\phi}\right )^{1/2}}=(-g_{tt})^{1/2}\,.
\nonumber
\end{equation}

The ZAMO would also conclude that at the Wald charge $\delta \varepsilon$ vanishes. But this argument is pretty weak, given its reliance on a particular observer. 

Furthermore, there is no value of $Q$ for which the covariant quantity $E\cdot B=0$ everywhere. In other words, although charges along the pole will not experience EM forces when the BH has the Wald charge, particles everywhere else will experience forces and will continue to flux around, creating regions of particle acceleration and therefore also the potential for EM radiation.
To  make unambiguous claims about the flow of charges requires we examine the dynamical equations, which we do next.

\section{Particle Acceleration}
\label{S:ParticleAcceleration}

Considering the equations of motion again,
\begin{equation}
\left ( p \cdot D\right ) p =  q F\cdot p.  
\label{Eq:LF}
\end{equation}
Notice that the Lorentz force, on the RHS, is proportional to the electric
field, $E_q=F\cdot u$, as perceived by the charged particle with 4-velocity
$u$ since in the particle's own frame there is no motion and so no magnetic
force.

Now, the Lorentz force does vanish on the poles at the Wald charge, but does not vanish off the poles. 
Furthermore, 
\begin{equation}
\frac{1}{4}Tr( F\cdot \tilde F)= E\cdot B \ne 0 
\end{equation}
off the poles so particles can slide along the field lines as we now show. 
Since $E\cdot B$ is a covariant quantity, we choose to examine $E\cdot B=E_Z\cdot B_Z$, in terms of the fields as measured by a ZAMO.
The electric field is
\begin{eqnarray}
E_Z 
&= & F\cdot u_Z \nonumber \\
&=& u_Z^t \left ( \partial A \cdot \left (\eta +\omegaz  \psi\right ) +  \left (\eta +\omegaz  \psi\right ) \cdot \partial A\right ).
\end{eqnarray}
The last term vanishes because of symmetries, giving
\begin{eqnarray}
E_Z 
&=& \partial A \cdot u_Z \nonumber \\
&=& -\partial V -  \partial u_Z \cdot A ,
\label{Eq:Ez}
\end{eqnarray}
where
\begin{equation}
V\equiv  -A \cdot u_Z 
\end{equation}
is the electrostatic potential as seen by a ZAMO.
At the Wald charge $V=0$ everywhere, but as Eq.\ (\ref{Eq:Ez}) shows, $E_Z$ is not necessarily zero everywhere when $V$ is.

Meanwhile, since $u_\phi=0$,
\begin{eqnarray}
B_Z &=& -\frac{1}{2 }\tilde \epsilon^{\mu \nu  \alpha \beta} F_{\alpha \beta} u_\nu
\nonumber \\
&=& -\frac{1}{2}\tilde \epsilon^{\mu t \alpha \phi} \partial_{\alpha}A_{\phi} u_t 
\end{eqnarray}
With $\tilde \epsilon^{t r\theta \phi}=\frac{1}{\sqrt{-g}}\epsilon_{t r \theta \phi}$ and the Levi-Civita symbol is defined by permutations of $\epsilon_{t r \theta \phi}=1$. Since $\tilde \epsilon^{\mu t \alpha \phi}=-\tilde \epsilon^{t\ i j\phi } $, we can write 
\begin{eqnarray}
E_Z\cdot B_Z 
&=& \frac{1}{2} \tilde \epsilon^{i j \phi} \left (\partial_{i}A \right )\cdot u_Z\left (\partial_{j}A_{\phi} \right )u_t  \nonumber \\
&=&
\frac{1}{2} 
\tilde \epsilon^{i j \phi} \left (\partial_{i}A_t \right )\left (\partial_{j}A_{\phi} \right )u_t u^t \nonumber \\
&=&
\frac{1}{2\sqrt{-g}} 
\partial_{[r}A_t \partial_{\theta ]}A_{\phi}  
= E\cdot B
\end{eqnarray}
using $u_tu^t=-1$. The final expression no longer depends on the velocity of the observer, which is gratifying. The expression is valid for all $Q$ and for all $\theta$ and actually only relies on the axisymmetry and stationarity of the spacetime and the vector potential.

\begin{figure*}
\begin{center}$
\begin{array}{c c}
\includegraphics[scale=0.4]{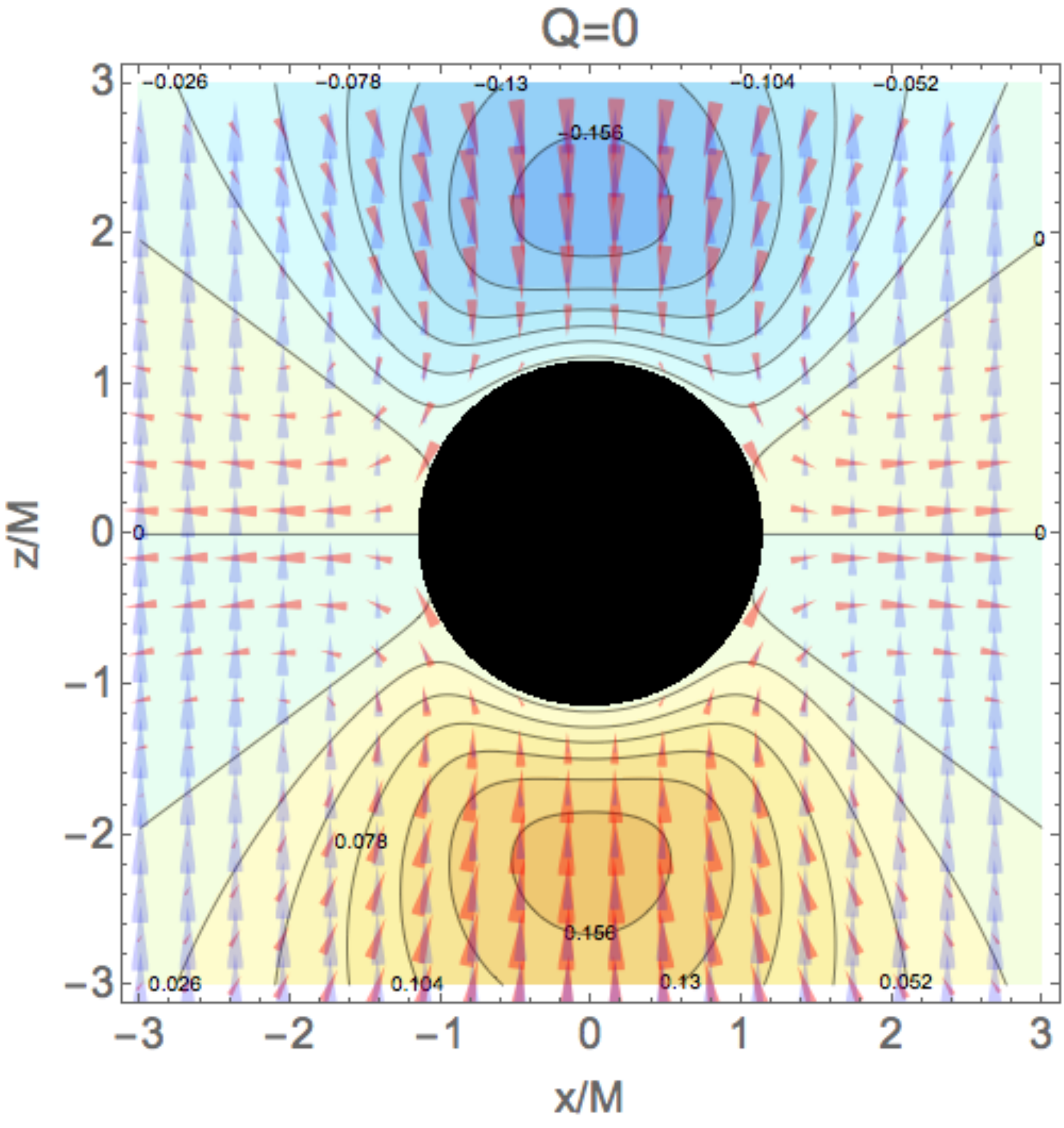} &
\includegraphics[scale=0.4]{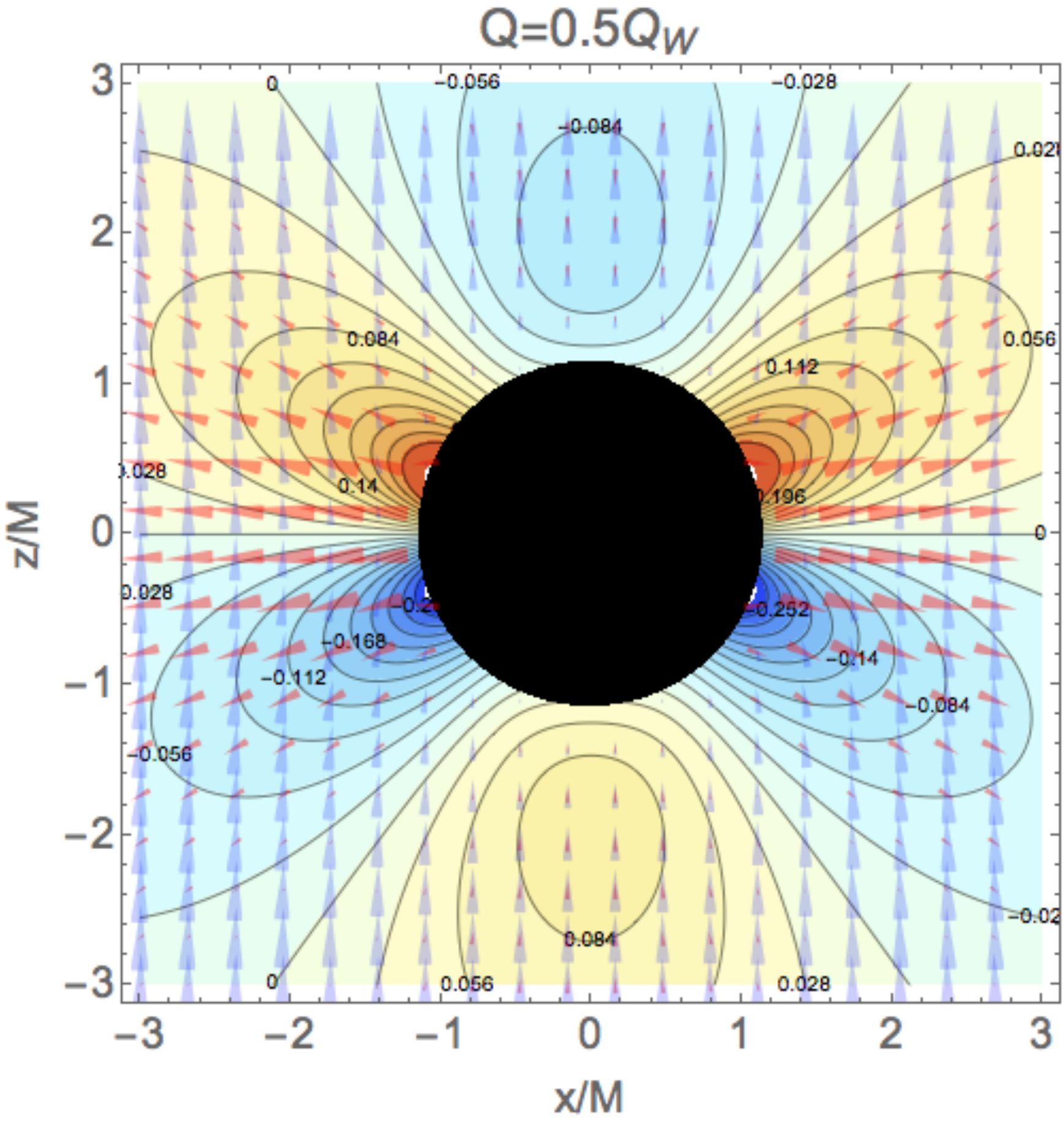} \\
\includegraphics[scale=0.4]{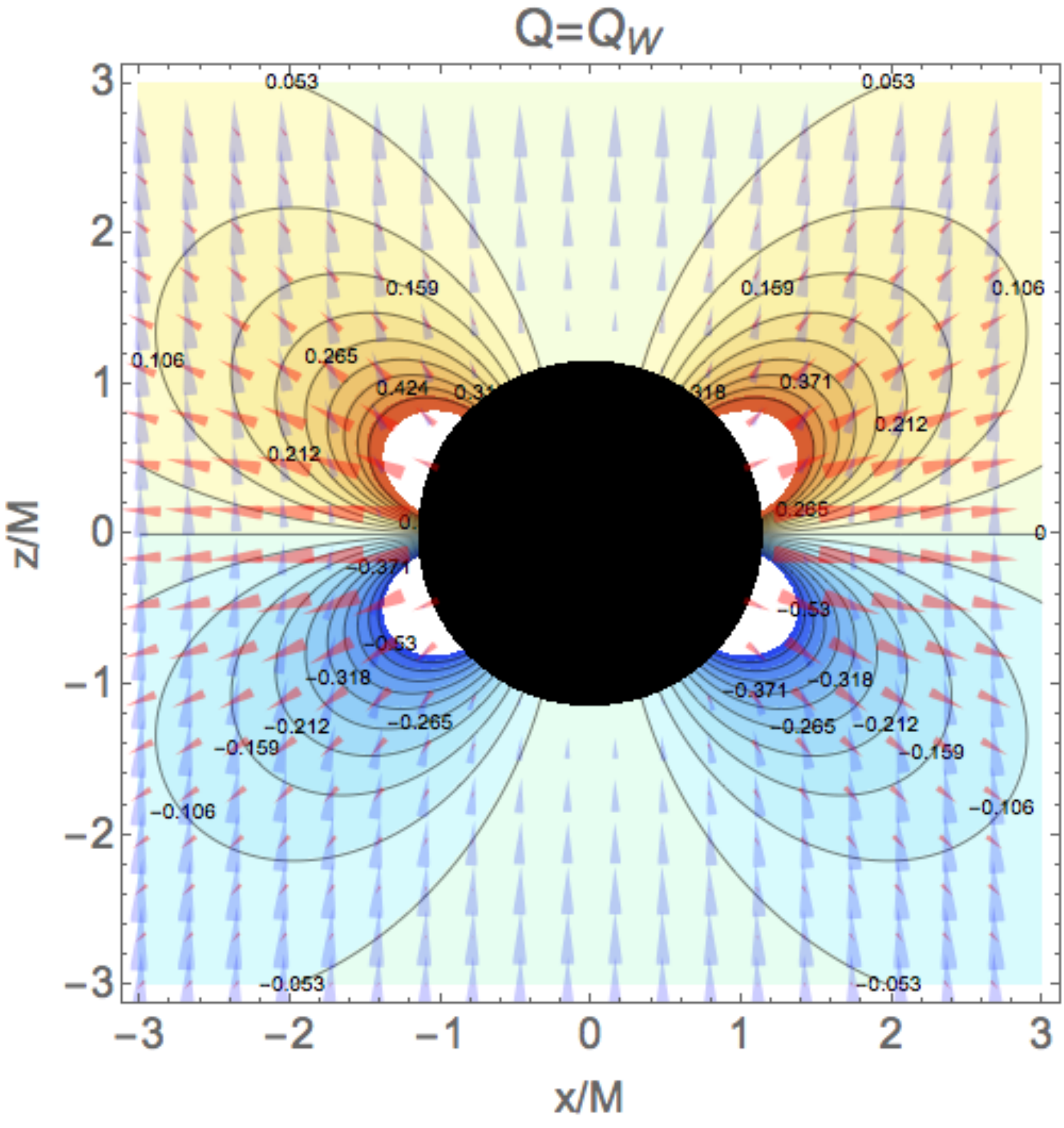} &
\includegraphics[scale=0.4]{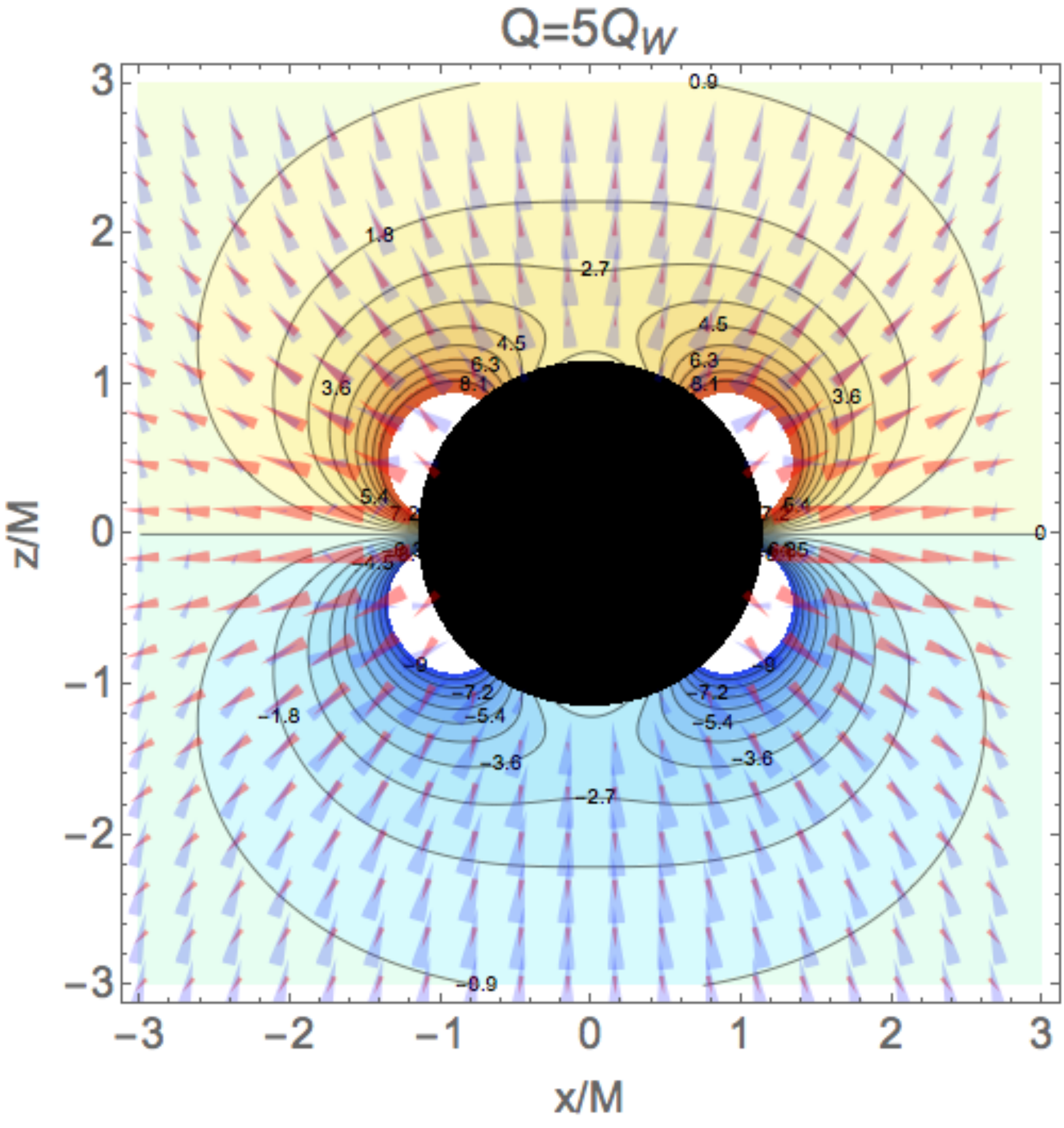} 
\end{array}$
\end{center}
\caption{
Shaded contours of $E \cdot B/B^2_o$ for the specified BH charge. Yellow shading is where $E \cdot B>0$ and blue shading represents where $E \cdot B<0$. Magnetic field vectors, as seen by ZAMOs, are drawn as blue triangles while ZAMO electric field vectors are drawn as red triangles. Each panel, from left to right, top to bottom is drawn for an increasing value of the BH charge Q, in units of the Wald charge, $Q_W$. The black sphere sphere represents the BH horizon.
}
\label{Fig:EdotB}
\end{figure*}

And here we get to the crux, there is no value of $Q$ for which $E\cdot B=0$
for all $\theta$ as evidenced by the sequence of plots in Fig.
\ref{Fig:EdotB}. This is also apparent explicitly on substitution of the
expressions for $E_Z$ and $B_Z$ in the above equation. At the Wald charge,
$E\cdot B=0$ at the poles and on the equator, but at no other values of
$\theta$.

Since the Wald charge cannot kill $E$ or $E\cdot B$, it must be that charges
are accelerated along the $B$-lines. At first we wondered if this suggested
that the charge on the BH is not stable. But investigating the flow of
charges in the following section reveals that the BH continues to absorb
positive and negative charges in equal measure, maintaining $Q=Q_W$.

\subsection{Orbits of test charges}

If we could analytically calculate a current density ${\J}$, then we could compute a flux across the horizon
\begin{equation}
\Phi \propto \oint_{\mathcal{H}} \J \cdot dA,
\end{equation} 
and determine if the flux is overall positive, negative, or zero. 

The current density is given by the product of the charge density and its four
velocity $\J = \rho \mathbf{u}$. For the purpose of computing the sign of the
horizon charge flux, we can follow single charges and compute the quantity $q
\mathbf{u}$, for test-charge $q$.  We won't be concerned about the EM fields due to
these test charges. So if we can generally solve for $u$, for any value of the
BH charge and initial conditions of the charged test particle, then  we can
compute the sign of the horizon charge flux and answer our question.

If we could determine an analogue to the Carter constant, we would also know $\pi^\theta = m u^\theta $.
Then we could use
\begin{equation}
u\cdot u = - 1 
\end{equation}
to solve for $u^r$. We would then know the current and the flux. However, as we've already argued we do not in general have a Carter constant and so we cannot calculate $u$ analytically in general.

We can easily calculate the 4-velocity and thereby the flux at the poles.
Start a particle at rest at $r_i$ along the poles
\begin{equation}
u_i=\sqrt{\frac{(r_i^2+a^2)}{\Delta_i}}\, \eta \, .
\end{equation}
To find the orbit from this initial condition we use the constant of motion $\varepsilon$ ($\ell=0$) evaluated at $\theta=0$. The energy is fixed by the initial conditions:
\begin{equation}
\varepsilon_i = 
m\left ( \frac{\Delta_i }{r_i^2+a^2}\right )^{1/2}+\frac{q}{2M}\left  (2B_o\a M -Q\right )\left ( \frac{\Delta_i }{r_i^2+a^2}\right )\nonumber.
\end{equation}
Since this energy is conserved, we can set $\varepsilon = -\left (mu^tg_{tt}+qA_t \right )$ equal to its initial value $\varepsilon_i$ to solve for $u^t$
\begin{equation}
mu^t=\frac{\varepsilon_i ( r^2+a^2)}{\Delta}-\frac{q}{2M}\left  (2B_o\a M -Q\right ).
\end{equation}
Notice that $u^t$ blows up at the horizon confirming infinite time dilation at the horizon.
Then from $u\cdot u=-1$, we have the remaining component of the 4-velocity
\begin{equation}
u^r=\pm \left ( -1 +\frac{\Delta }{r^2+a^2}(u^t)^2 \right )^{1/2}\left (\frac{\Delta}{\Sigma}\right )^{1/2}.
\end{equation}

At the poles the current into the event horizon, which has a radial normal, is just
$\J=qu$ and clearly this current is independent of charge only at the Wald value of $Q$ because $\varepsilon$ becomes independent of $q$, as does $u^t$ and therefore $u^r$.

To find the charge flux across the horizon anywhere else, we numerically
integrate the orbits of oppositely charged particles.

As initial data we are free to set the clock to $\tau=0$ and the initial
$\phi(0)=0$ due to the symmetry of the metric. We choose to start orbits at
rest $u^r_i=u^\theta_i=u^\phi_i=0$ at the radius $r_i=40M$ (unless specified). We then vary the
initial $\theta_i$ over the range $0 \le \theta_i \le \pi/2$. The timelike
condition $u\cdot u =-1$ fixes $u_i^t=\sqrt{-1/g_{tt}}$. The energy and
angular momentum are then found from Eq. (\ref{Eq:EL}), and will depend on
$\theta_i$ through the metric components:
\begin{eqnarray}
\varepsilon &=  m\sqrt{-g_{tt}} - q A_t  \nonumber \\
\ell &=  m\frac{g_{t\phi}}{\sqrt{-g_{tt}}} + q A_{\phi} .
\end{eqnarray}
Notice that each orbit has a different $\varepsilon$ and $\ell$ depending on
its initial $\theta_i$ and charge.

For each initial value of $\theta_i$ we numerically compute the trajectories
of a test mass with positive and negative charge. While we do not extensively
explore all possible initial conditions, we note that the at-rest initial
condition orbits considered here have a similar quality in that they all orbit
at a constant cylindrical radius and rotate azimuthally around the BH.  The
charges stay very close to their starting cylindrical radius because they are
confined to the vertical magnetic field lines. The azimuthal rotation is due
to the $\mathbf{E} \times \mathbf{B}$ drift, in the same direction for both
signs of test charge.

\begin{figure*}
\begin{center}$
\begin{array}{c}
\includegraphics[scale=0.33]{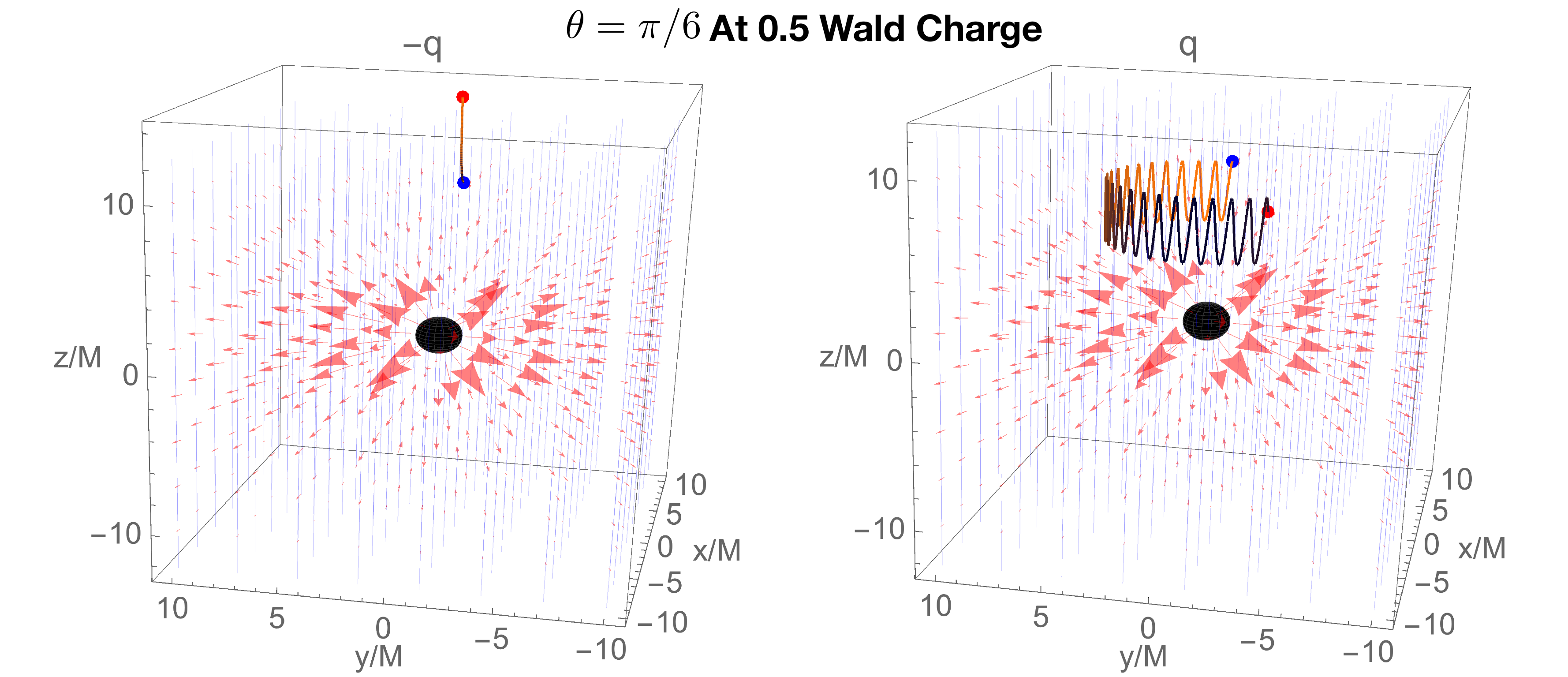} \\
\includegraphics[scale=0.33]{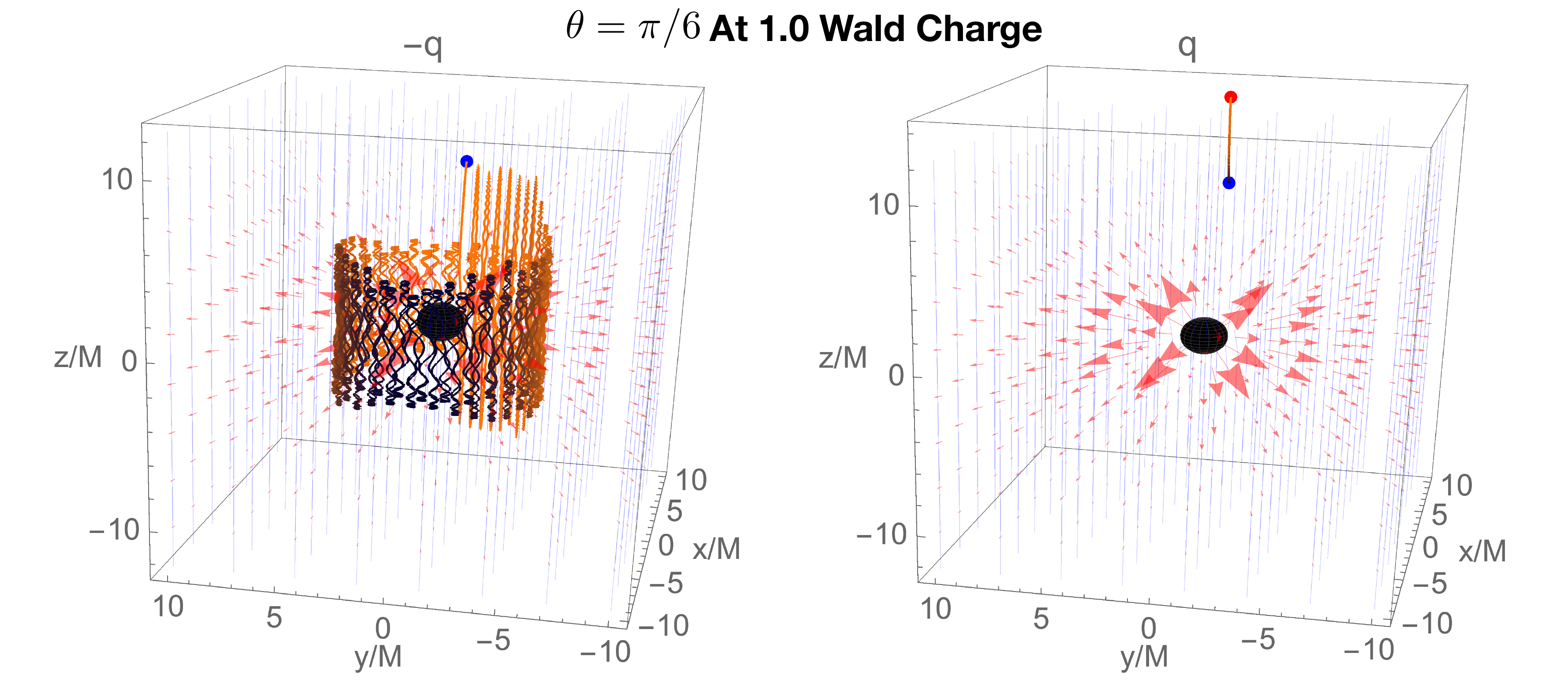} \\
\includegraphics[scale=0.33]{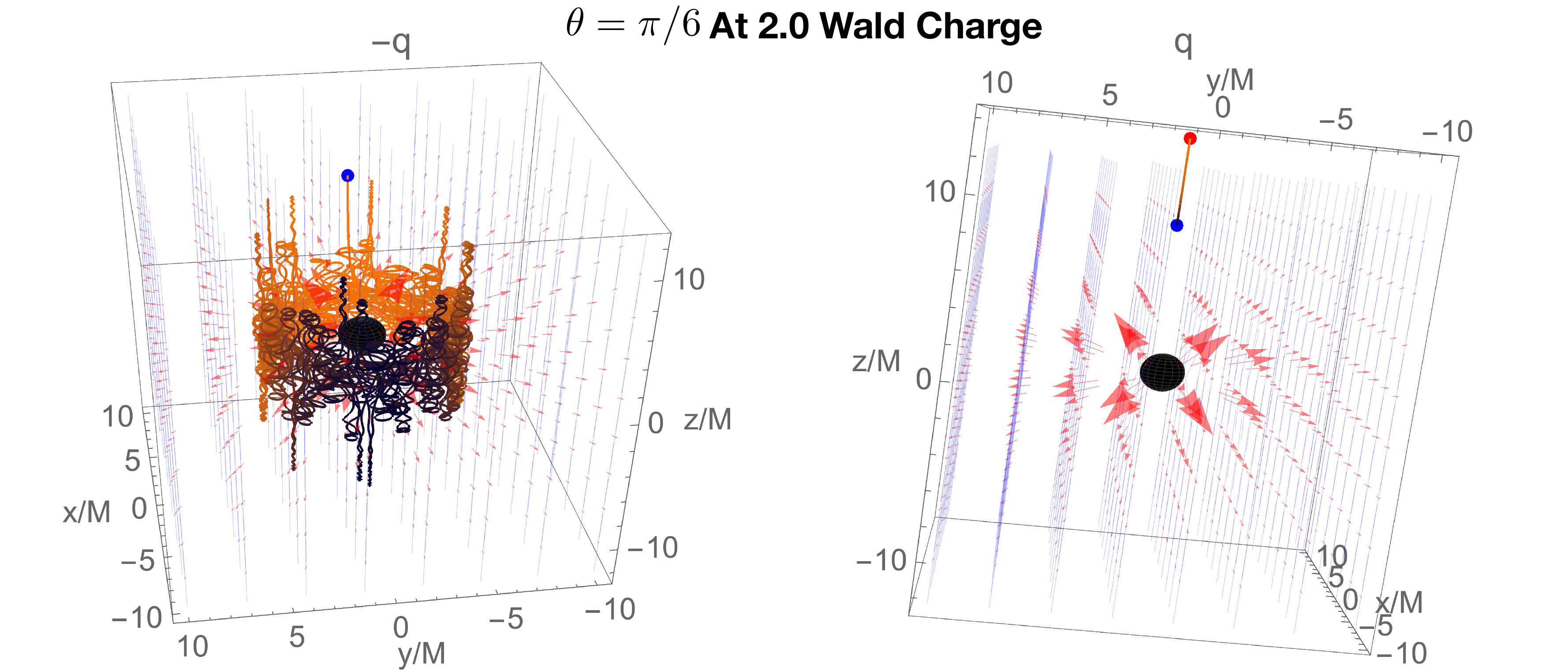}
\end{array}$
\end{center}
\caption{
Examples of orbits of charged particles around the spinning BH in the Wald
field, with at-rest initial conditions ($r_i=10M$, $\theta_i=\pi/6$). Red
triangles are electric field vectors and blue lines represent the uniform
immersing magnetic field, aligned with the BH spin axis. The left column is
for negatively charged test charges and the right column is for positively
charged test charges. The initial position of the charge is marked by a blue
dot and the final position is marked by a red dot.
}
\label{Fig:Orbits}
\end{figure*}

\begin{figure}
\begin{center}$
\begin{array}{c}
\includegraphics[scale=0.4]{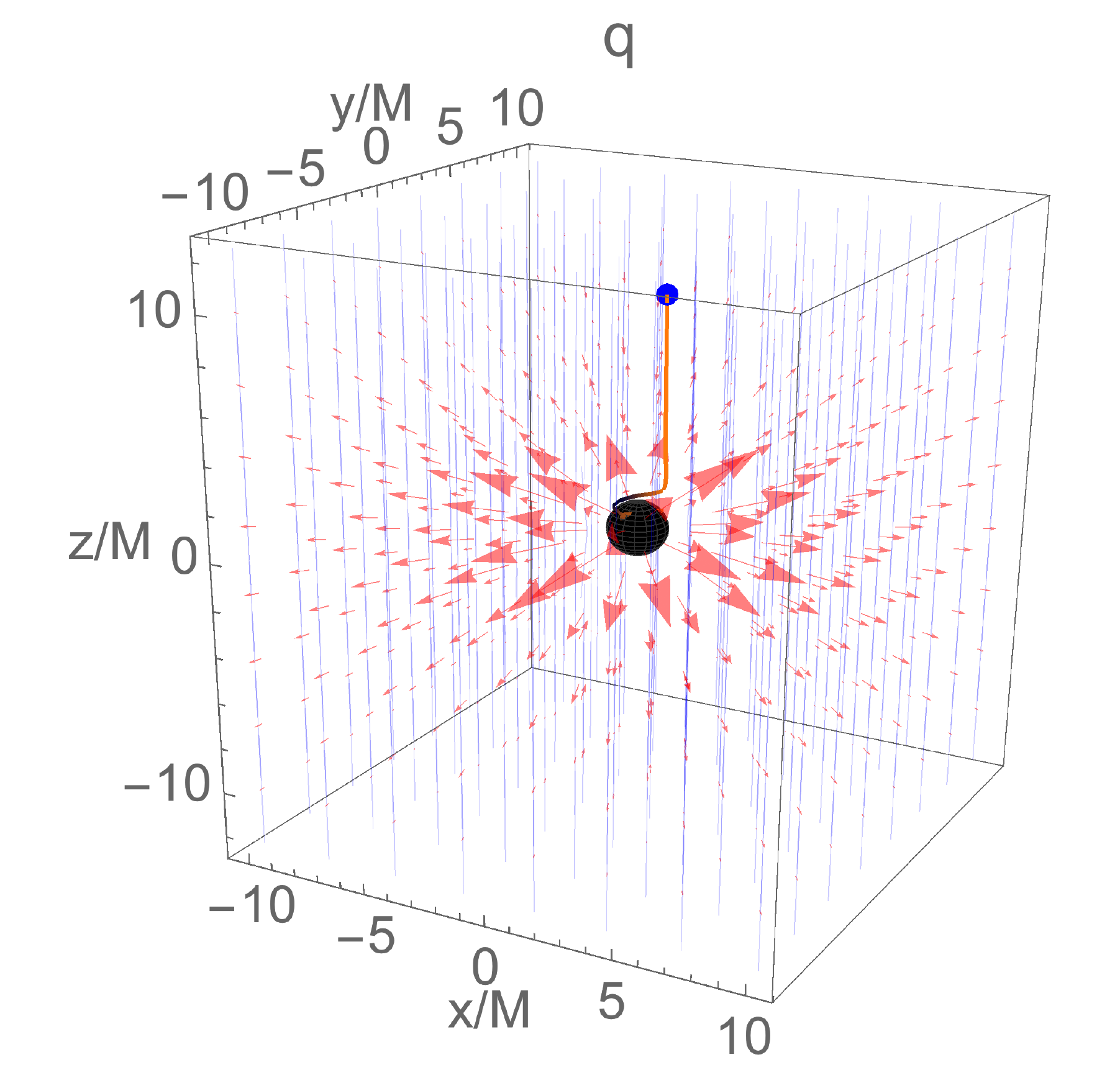} 
\end{array}$
\end{center}
\caption{
The same as Figure \ref{Fig:Orbits} but for a plunging orbit.
}
\label{Fig:OrbitPlunge}
\end{figure}

This simple qualitative behavior can lead to a number of different fates for the test charge for which we plot examples in Figure \ref{Fig:Orbits}, and categorize into four types:
\begin{itemize}
\item Expulsion from the system along B-field lines, when $\mathbf{E} \cdot \mathbf{B}$ is initially directed out of the system for the given test charge sign (see the top-left panel of Figure \ref{Fig:Orbits}).
\item Plunge into the horizon for charges that start at small values of $\theta_i$, such that their initial cylindrical radius $r_i \sin \theta_i$ is small (see Figure \ref{Fig:OrbitPlunge}).
\item Regular vertical oscillations (in the direction of the magnetic field and BH spin axis) at fixed cylindrical radius for BHs with charge below the Wald charge (see the top-right panel of Figure \ref{Fig:Orbits}).
\item Non-regular vertical oscillations at fixed cylindrical radius for BHs with charge above the Wald charge (see the bottom-left panel of Figure \ref{Fig:Orbits}).
\end{itemize}

These orbit types can be understood from the electric and magnetic field
structure of the Wald solution for different BH charges. The electric and
magnetic field vectors, along with contours of $\mathbf{E} \cdot \mathbf{B}$
are plotted in Figure \ref{Fig:EdotB}. The primary change in field
configuration with increasing BH charge is the dominance of a quadrupolar
electric field below the Wald charge vs. a predominately monopolar electric
field above the Wald charge. This follows since the electric field sourced by the
monopole of charge on the BH eventually dominates over the quadrupolar
electric field generated by a Kerr BH in a uniform magnetic field
\citep[\textit{}][]{MPBook}. The transition occurs at the Wald
charge, at which point the electric field at the poles becomes zero, having
opposite sign in the z direction (direction of BH spin) below and above the
Wald charge. Across all cases the x-component (direction perpendicular to BH
spin) of the electric field does not change appreciably.

For each of the panels for which $Q\neq Q_W$ in Figure \ref{Fig:EdotB}, it is
clear that there is a non-zero value of $\mathbf{E} \cdot \mathbf{B}$ above
the poles of the BH. This means that in the top left panel, for example,
positive charges that start at an initial position within the cylinder
containing the BH horizon ($x^2 + y^2 \lesssim r^2_+$) will follow a
trajectory directly into the BH (\textit{e.g.}, Figure \ref{Fig:OrbitPlunge}).
Test charges of the opposite sign of charge will be expelled from the BH
(\textit{e.g.}, the top left panel of Figure \ref{Fig:Orbits}).

Consider further the $Q=0$ case displayed in the top left panel of Figure
\ref{Fig:EdotB}. Moving farther in the x-direction from the BH (a larger
cylindrical radius), the charges with negative charge are still expelled as
long as they are in the blue-shaded region of negative $\mathbf{E} \cdot
\mathbf{B}$ in the top hemisphere, or in the yellow-shaded region of positive
$\mathbf{E} \cdot \mathbf{B}$ in the bottom hemisphere, where the E-field is
aligned to accelerate negative charges out of the system along the z-directed
B-field. The positive charges, however, are no longer guided by the B-field
into the BH horizon, rather they move in the negative z-direction until
crossing a line where $\mathbf{E} \cdot \mathbf{B}$ changes sign, and hence
the direction of the z-component of the E-field changes sign. This results in
a vertical oscillation of the test charge about the $\mathbf{E} \cdot
\mathbf{B}=0$ line at $\sim \pm 50^{o}$ in the top hemisphere of the $Q=0$
panel (and also the analogue in the lower hemisphere). $\mathbf{E}\times
\mathbf{B}$ drift causes the orbit to rotate azimuthally. This type of
regularly-oscillating orbit can be seen in the top right panel of Figure
\ref{Fig:Orbits}.

A similar vertical oscillation occurs around the equator ($\theta=\pi/2$) for
negative charges. To see why this is, again consider the upper hemisphere of
the BH magnetosphere in the $Q=0$ panel of Figure \ref{Fig:EdotB}. Negative
charges with initial conditions below the line where $\mathbf{E} \cdot
\mathbf{B}$ becomes positive will be forced downwards initially along magnetic
field lines until they cross the equatorial plane, where the  $\mathbf{E} \cdot
\mathbf{B}$ reverses sign again, forcing the negative charge back into the upper
hemisphere. Hence negative charges in the equatorial regions are not expelled.

A similar situation as described for the $Q=0$ case holds for $Q<Q_W$
(\textit{e.g.}, the top right panel of Figure \ref{Fig:EdotB}). A difference
being that the monopolar E-field sourced by $Q$ is added to the quadrupolar
E-field of the $Q=0$ case and causes $\mathbf{E} \cdot \mathbf{B}$ to change
sign at a smaller $\theta$ than for $Q=0$. This causes the region of stably
orbiting positive charges in the region between the poles and the equatorial
plane to shrink and move to higher latitudes until at the Wald charge this
region disappears because at $Q=Q_W$, $\mathbf{E} \cdot \mathbf{B}$ changes
sign only at the equator and goes to zero at the poles. Hence, at the Wald
charge, test charges of both signs fall in at the poles while in the
equatorial region negatively charged test charges orbit stably around the
positively charged BH, tracing out cylinders oriented along the z-axis.

For $Q>Q_W$, as illustrated in the bottom right panel of Figure
\ref{Fig:EdotB}, the E-field is dominated by a monopole resulting in an
$\mathbf{E} \cdot \mathbf{B}$ at the poles that is oppositely directed from
the $Q<Q_W$ case. Hence, for $Q>Q_W$ the BH prefers to discharge back to
the Wald charge along the poles. In the equatorial region there still exist
negative charges on vertically oscillating orbits of nearly constant
cylindrical radius. However, as shown in the bottom and middle left panels of
Figure \ref{Fig:Orbits}, the vertical oscillations occur in much more
complicated patterns than in the $Q>Q_W$ case. We leave investigation of these
orbits for future work.

\begin{figure}
\begin{center}$
\begin{array}{c c}
\includegraphics[scale=0.23]{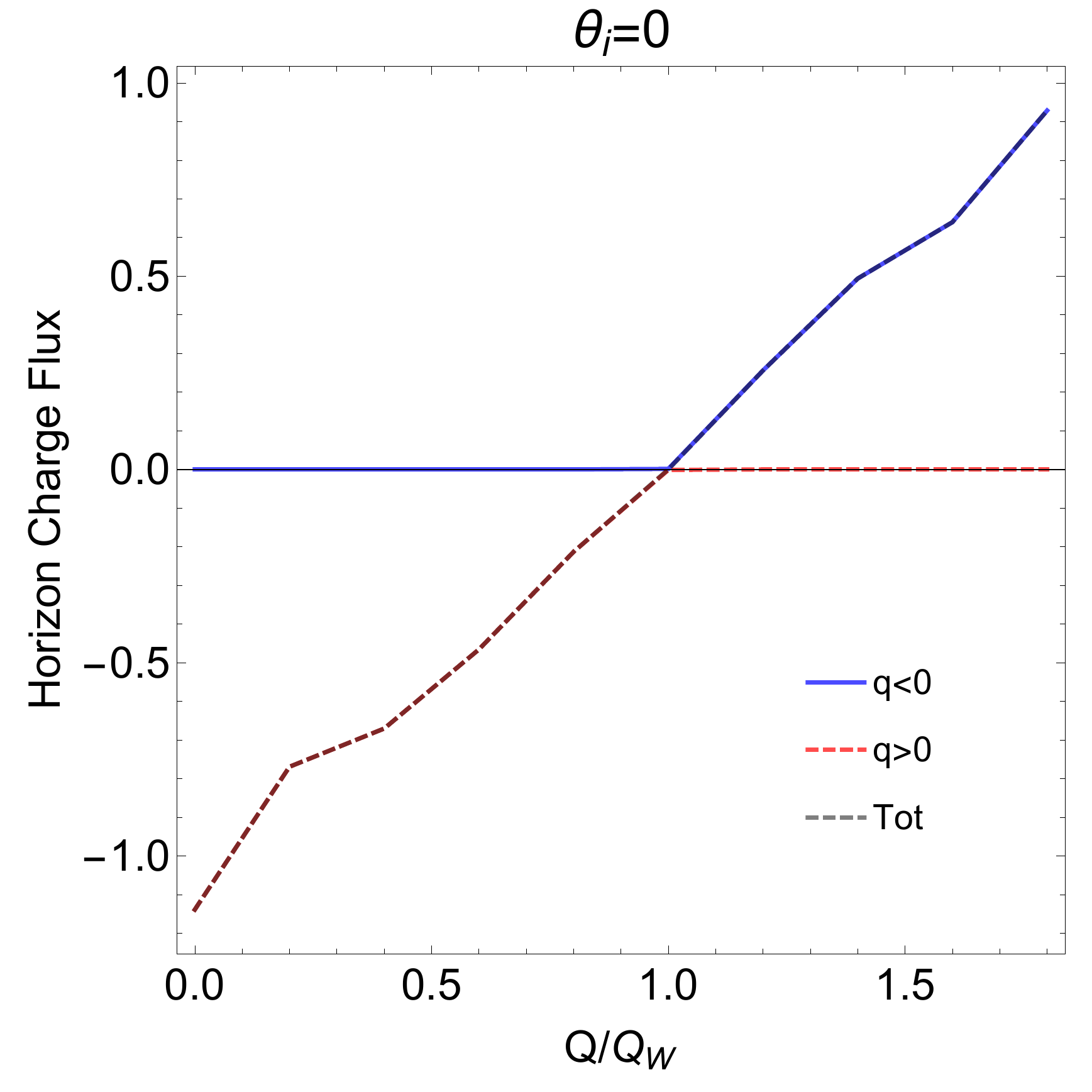} &
\includegraphics[scale=0.23]{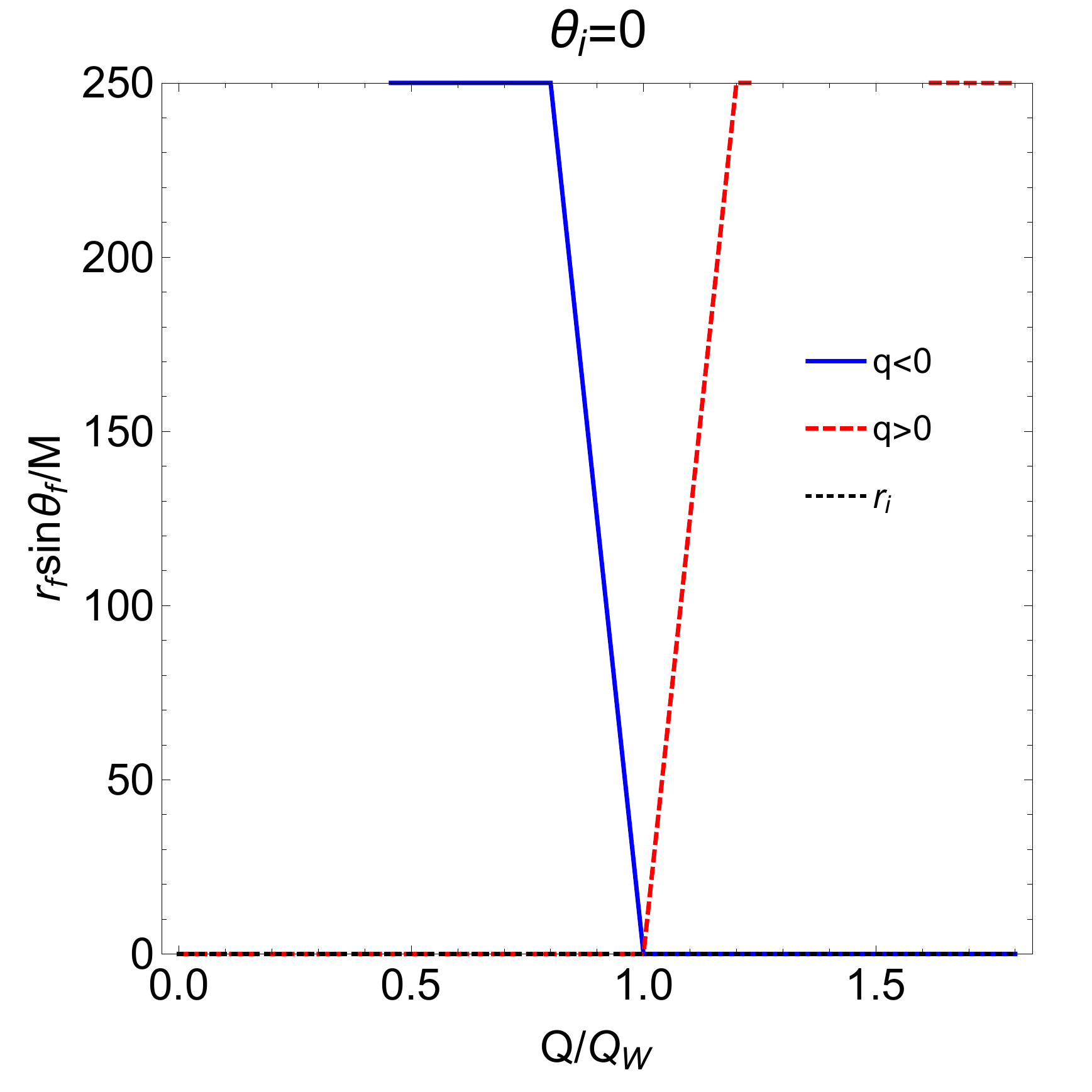} \\
\includegraphics[scale=0.23]{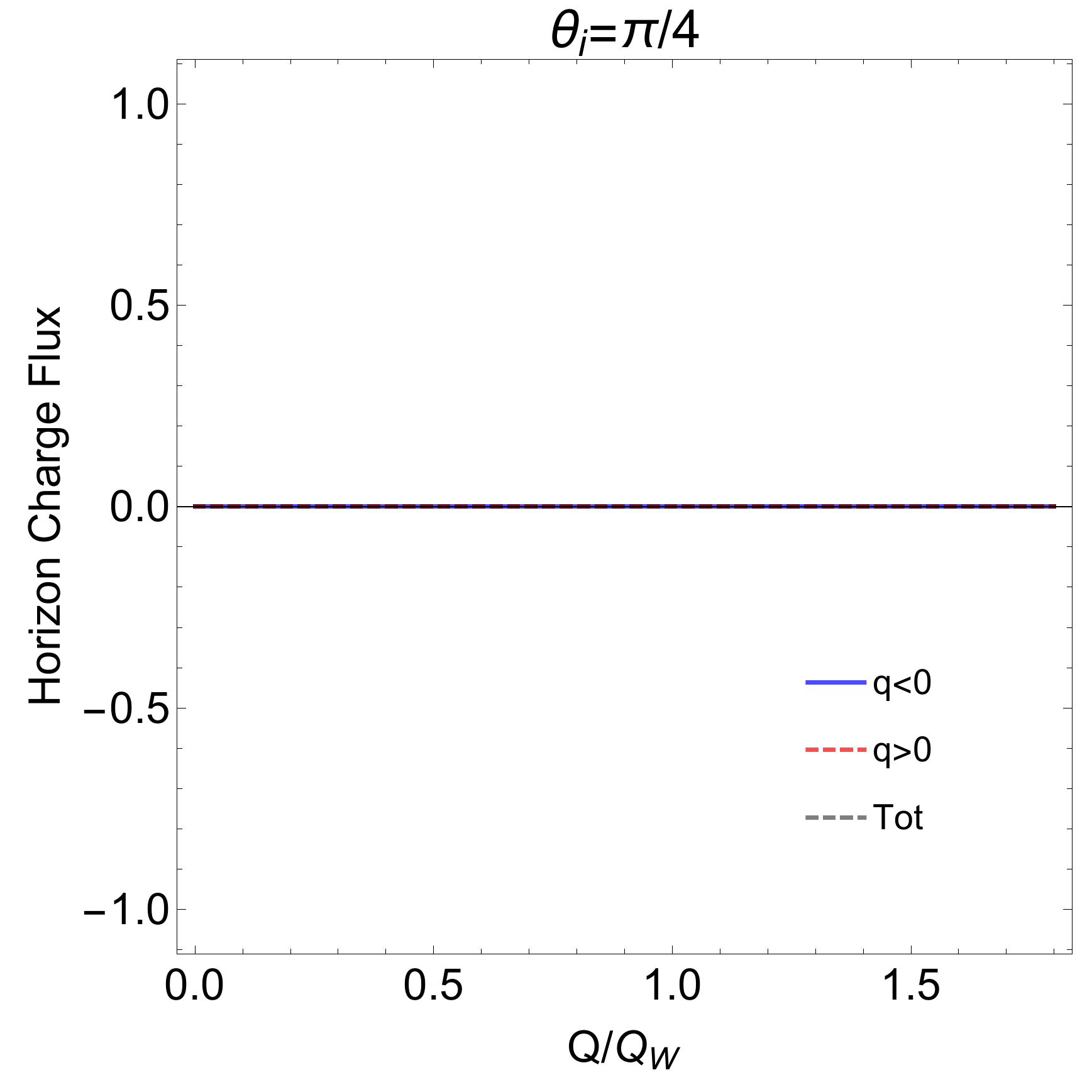} &

\includegraphics[scale=0.23]{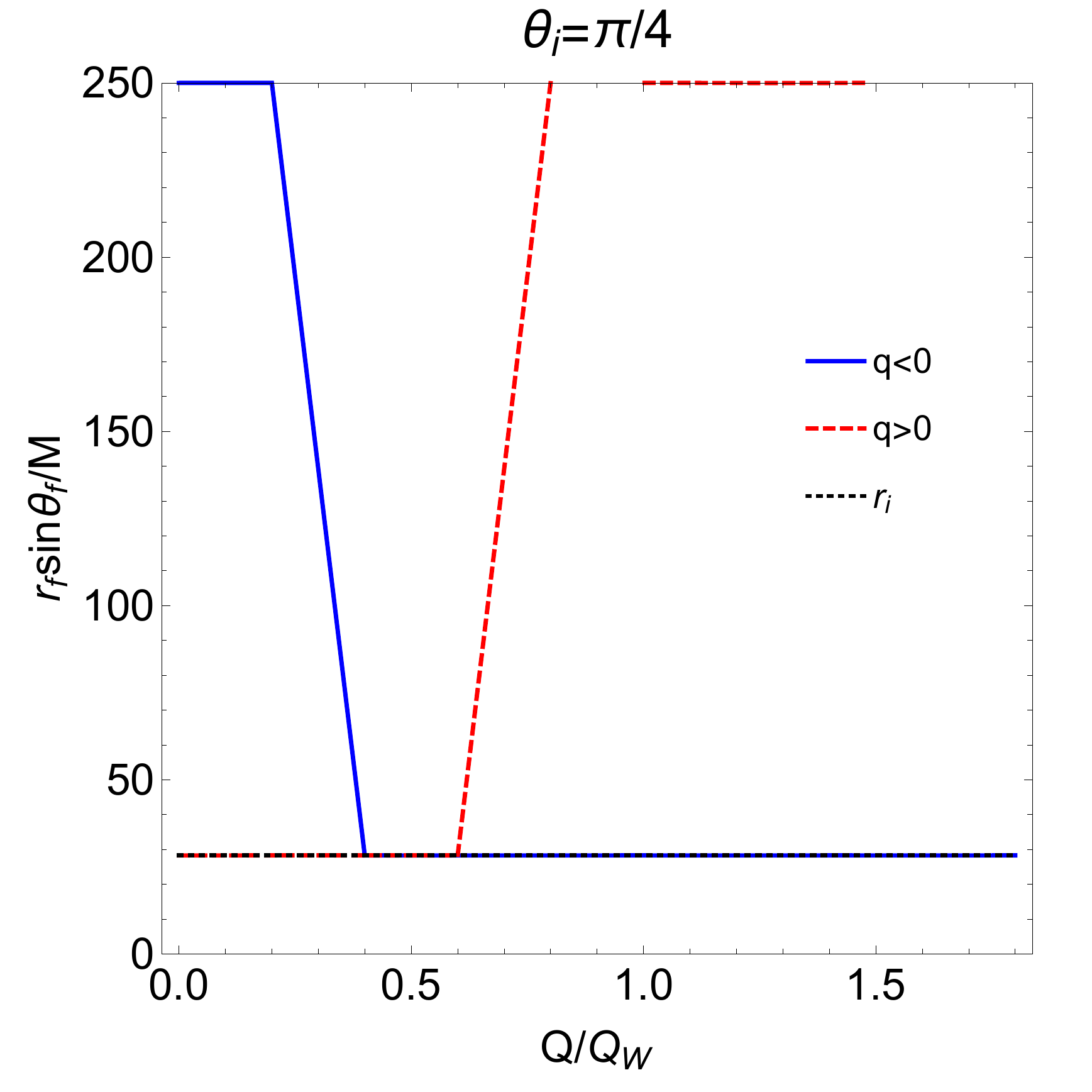} \\
\includegraphics[scale=0.23]{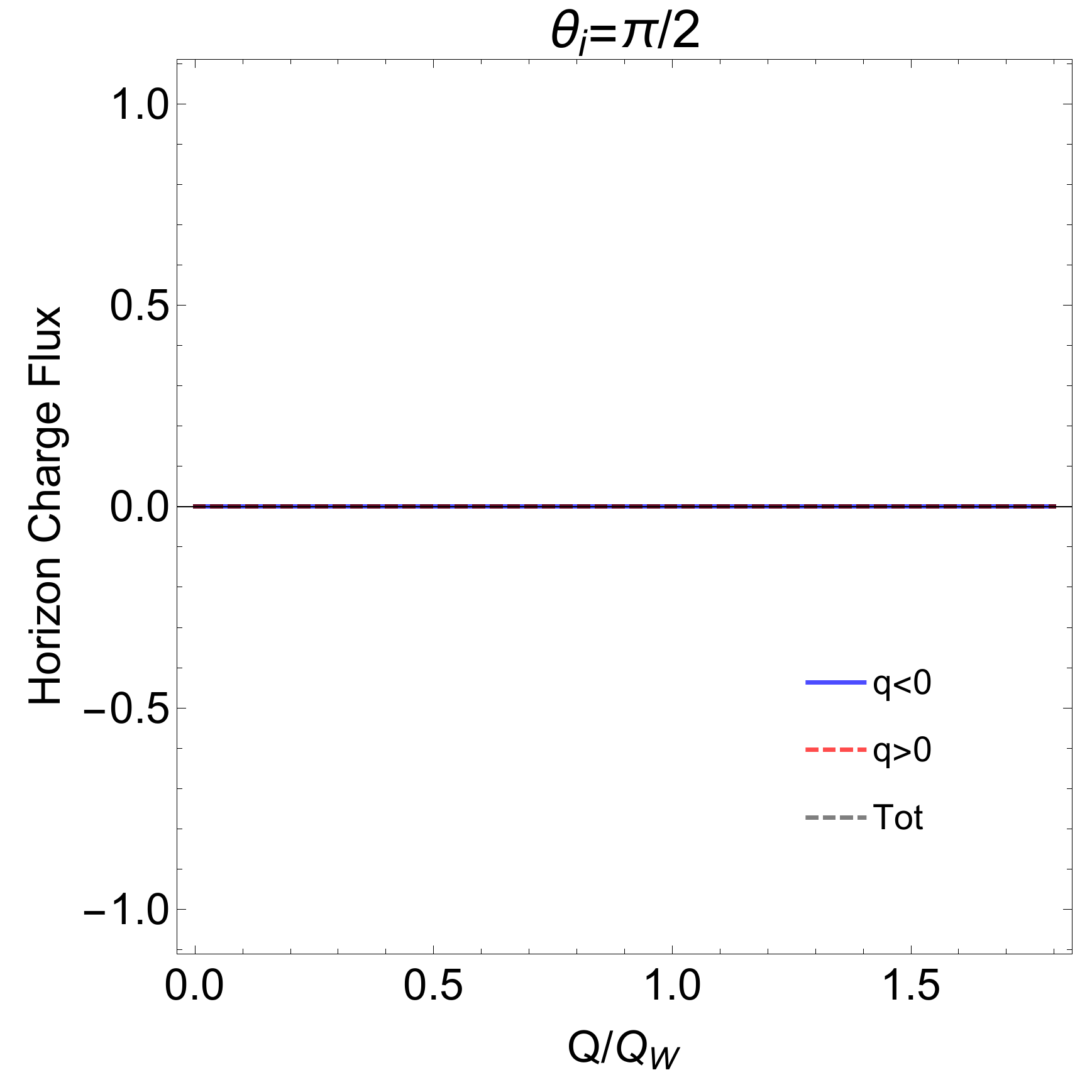} &

\includegraphics[scale=0.23]{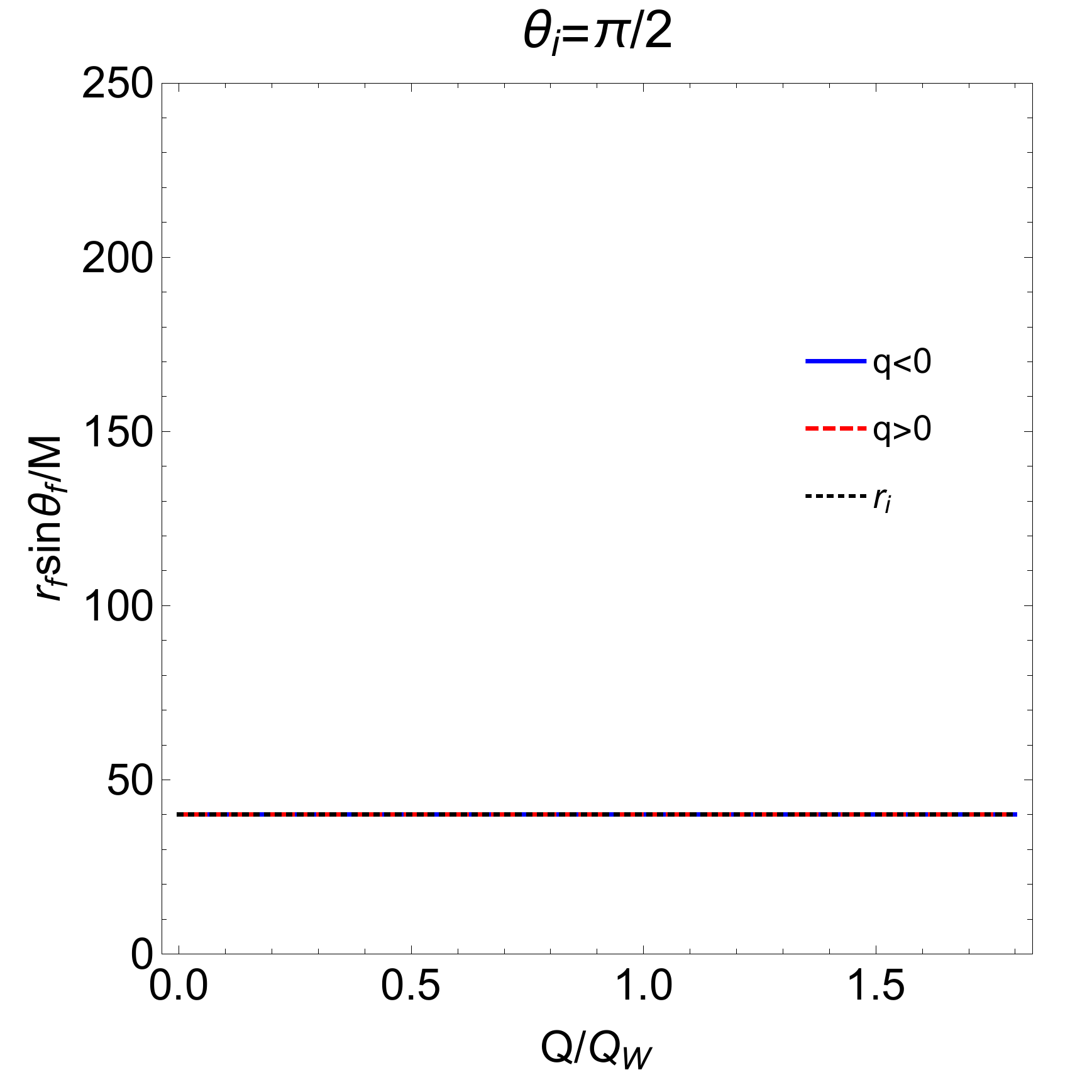}
\end{array}$
\end{center}
\caption{
The horizon charge flux (left column) and the final cylindrical radius of a test charge (right column) as a function of BH charge in units of the Wald charge, for the labeled initial theta coordinate, $\theta_i$ of the test charge. When a charge reaches spherical radius $250M$ we plot the final radius at this maximum value to show that it has been expelled.
}
\label{Fig:FvQ}
\end{figure}

\begin{figure}
\begin{center}$
\begin{array}{c c}
\includegraphics[scale=0.23]{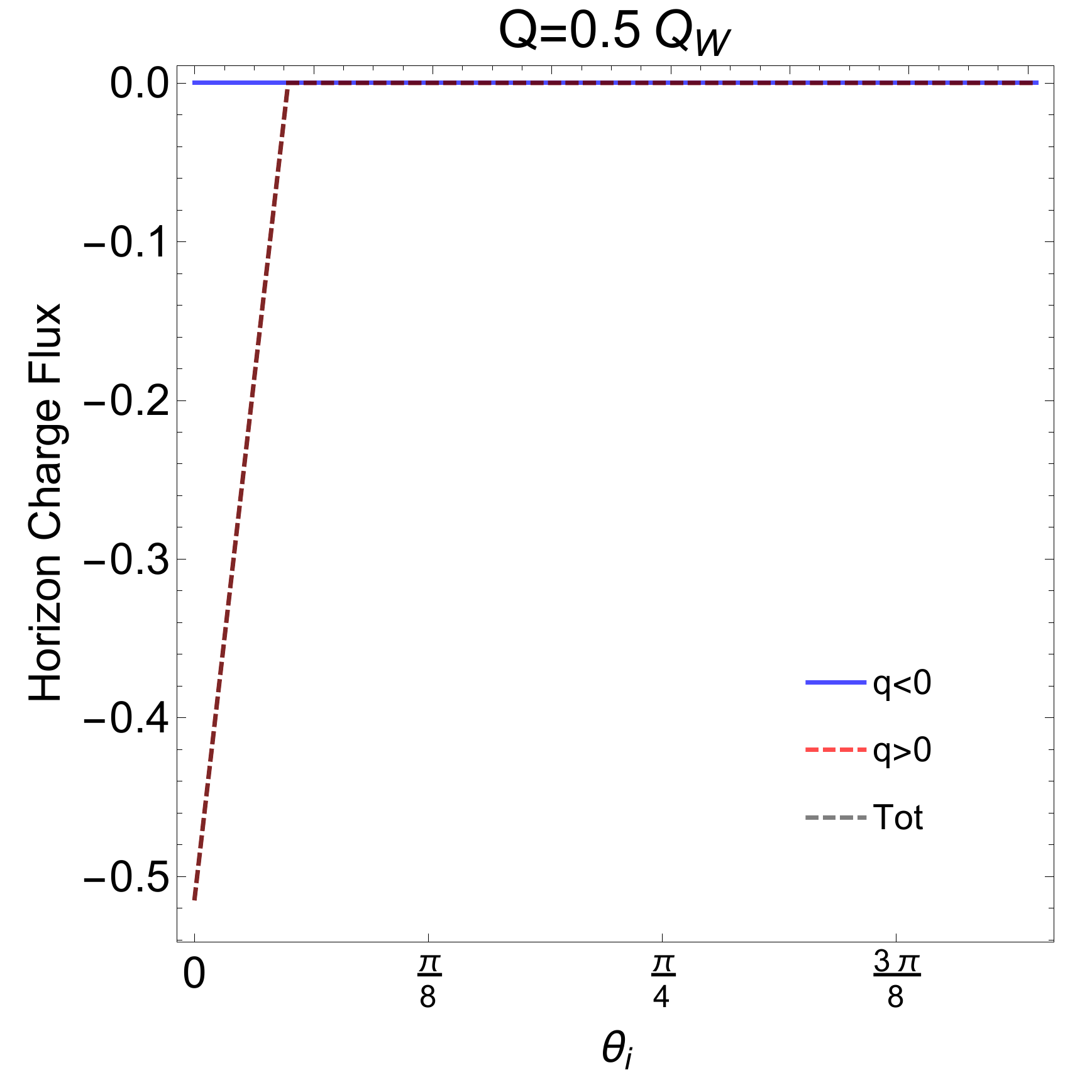} &
\includegraphics[scale=0.23]{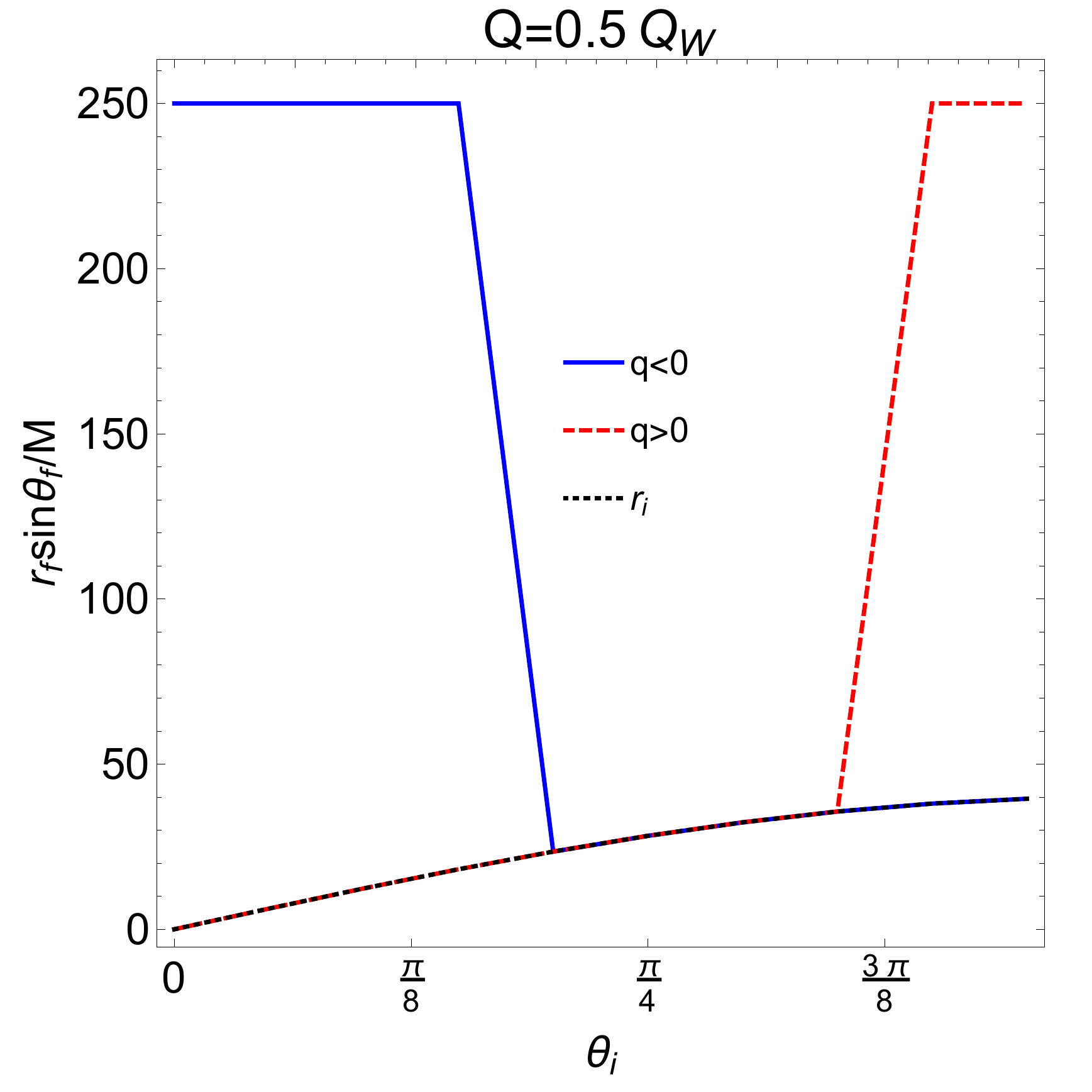} \\
\includegraphics[scale=0.23]{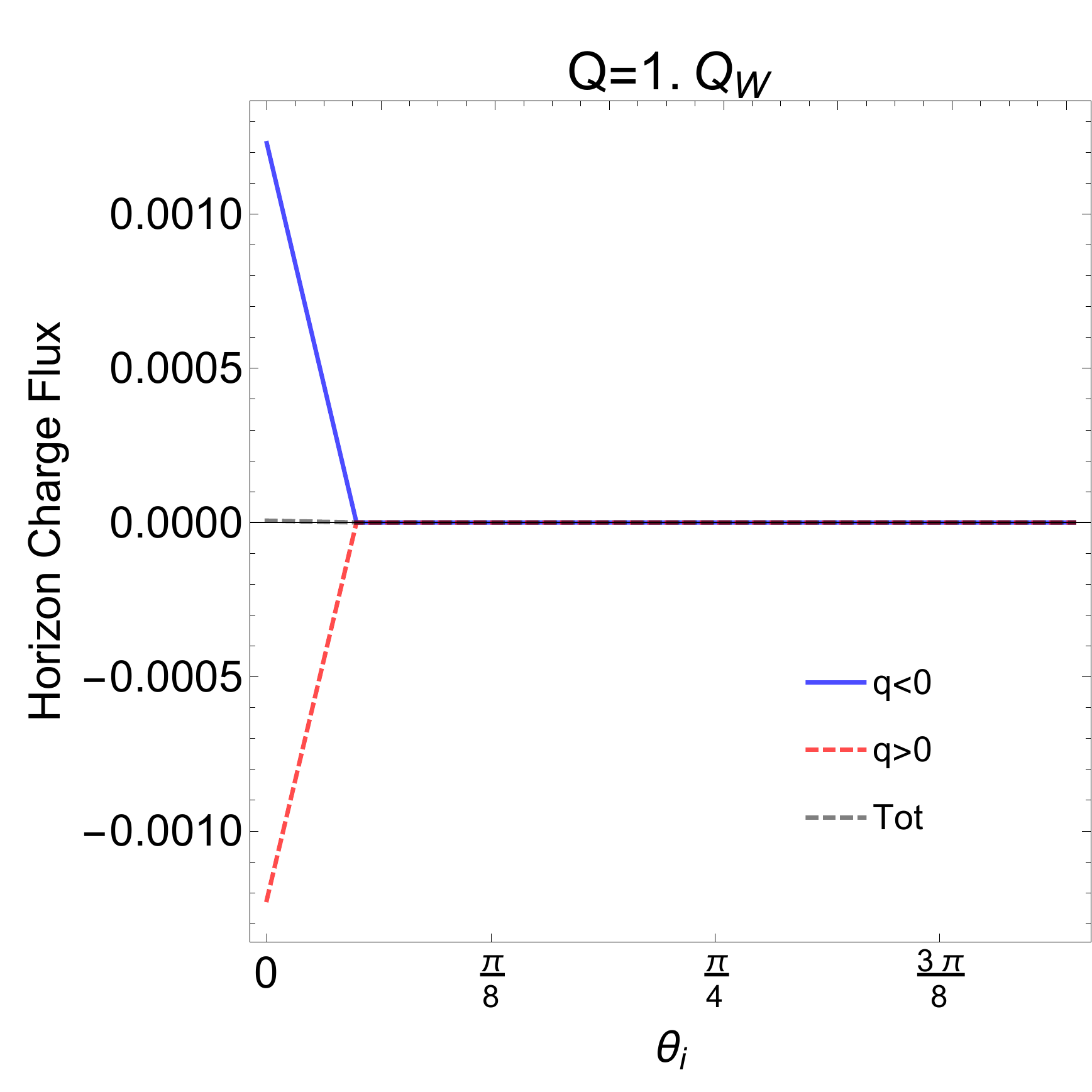} &
\includegraphics[scale=0.23]{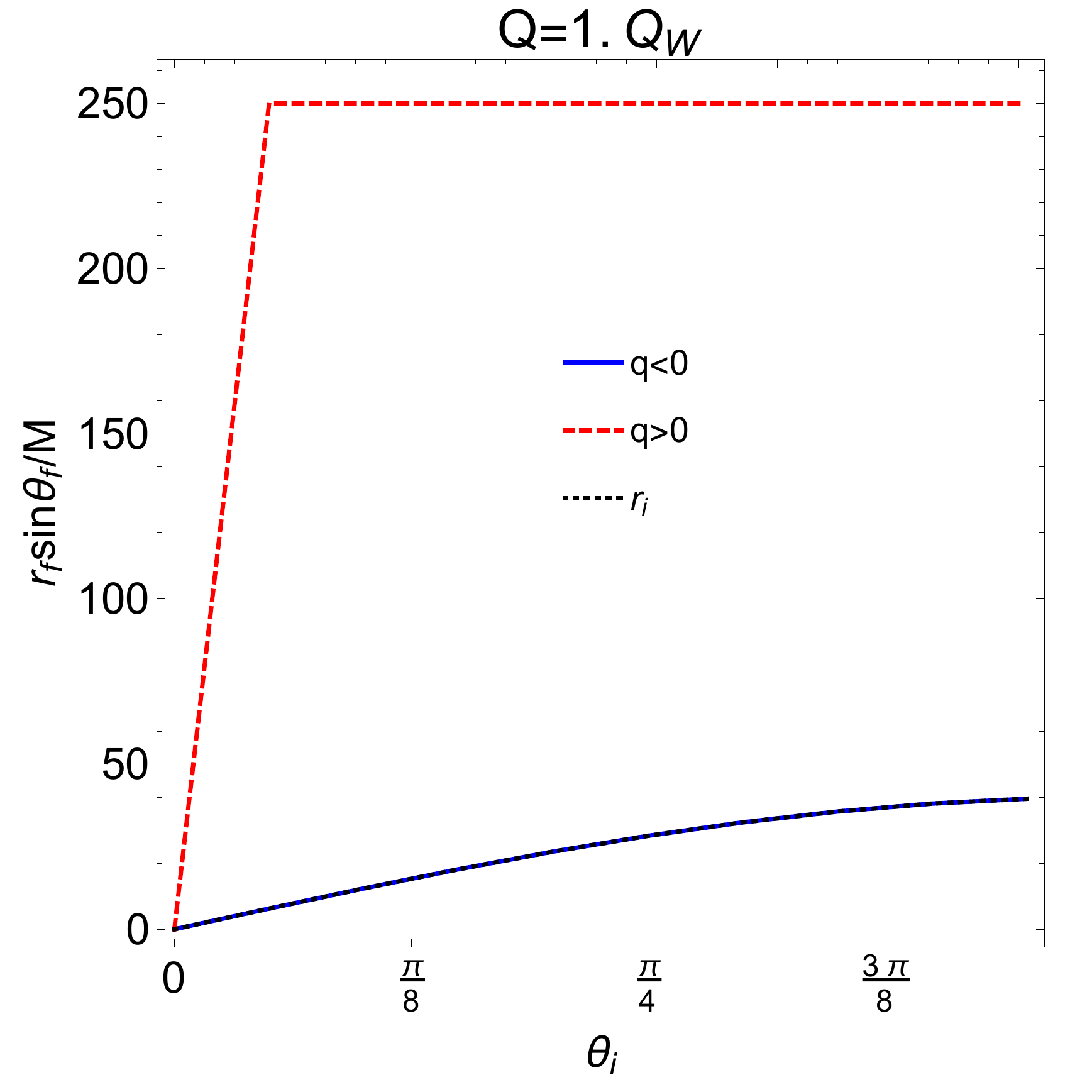} \\
\includegraphics[scale=0.23]{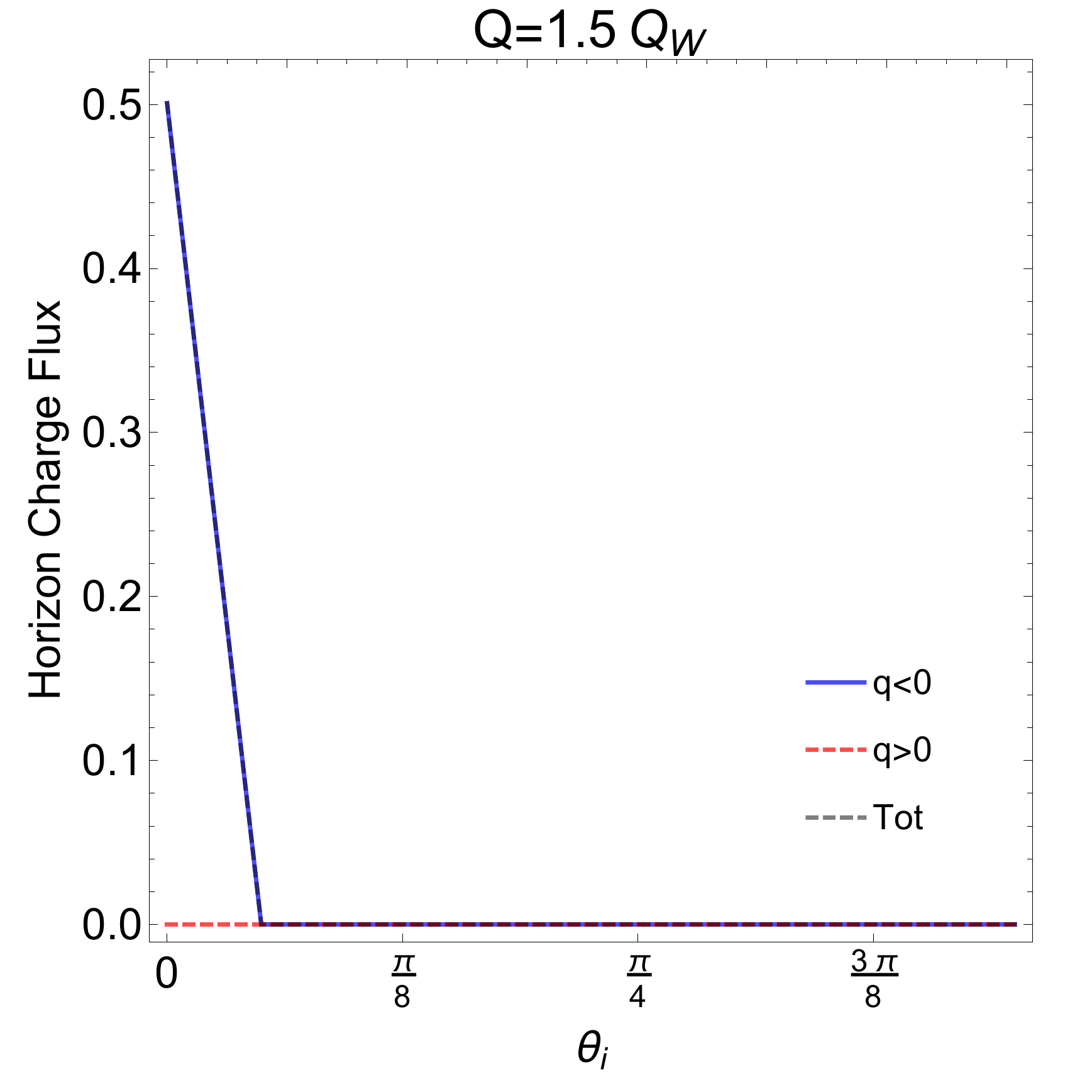} &
\includegraphics[scale=0.23]{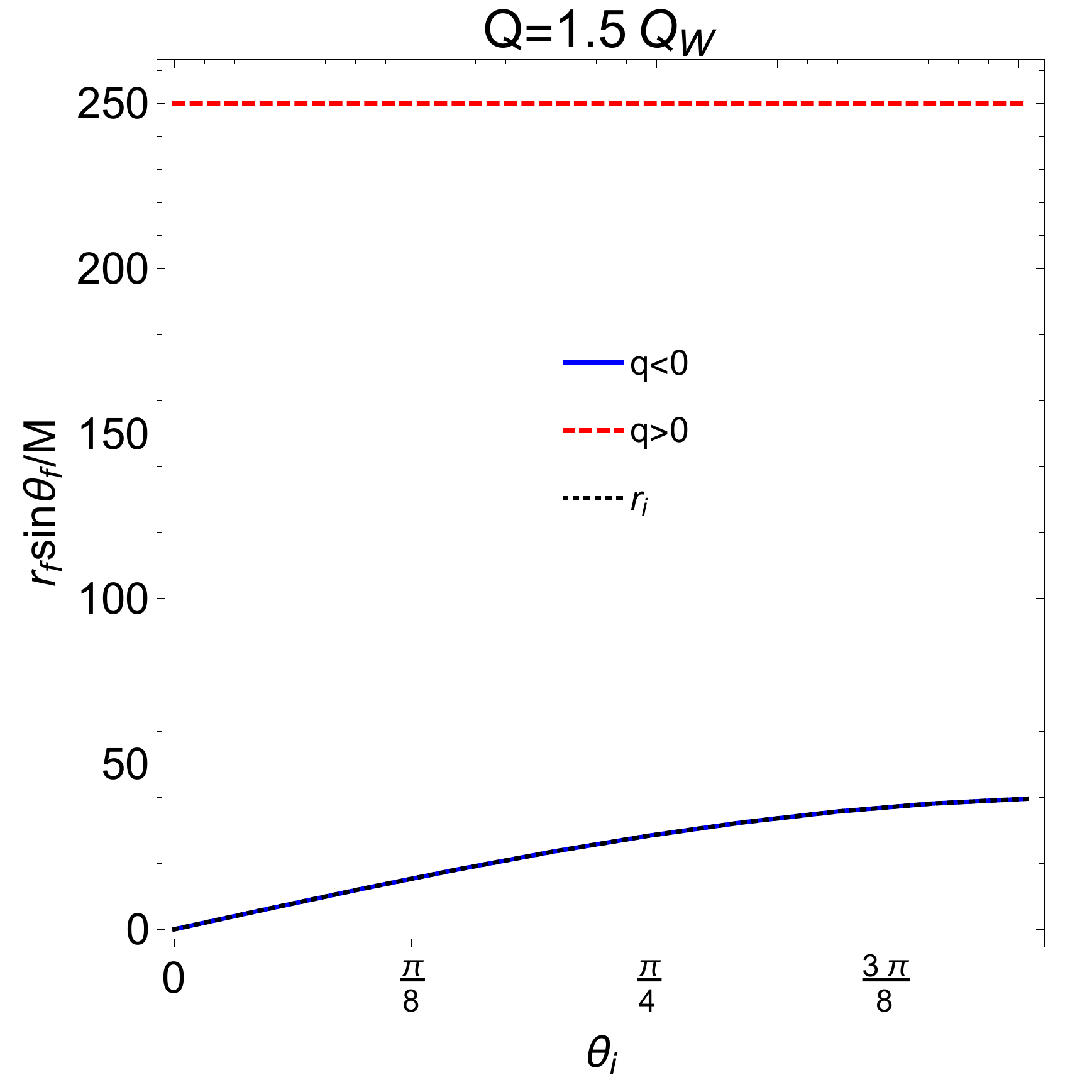}
\end{array}$
\end{center}
\caption{
The horizon charge flux (left column) and the final cylindrical radius of a test charge (right column) as a function of $\theta_i$ for a BH with the labeled charge in units of the Wald charge. When a charge reaches spherical radius $250M$ we plot the final radius at this maximum value to show that it has been expelled.
}
\label{Fig:Fvth}
\end{figure}

This general description of orbits in the Wald field leads to a global picture
of charging and discharging of a spinning BH in a uniform B-field that we
summarize with Figures \ref{Fig:FvQ} and \ref{Fig:Fvth}. In Figures
\ref{Fig:FvQ} and \ref{Fig:Fvth} we display the fate of test charges for grids of
initial conditions. In the left columns of Figure \ref{Fig:FvQ} and
\ref{Fig:Fvth}, we display the total flux evaluated at the horizon for
positively charged particles (red-dashed line), negatively charged particles
(blue line), and the total from both charges (black-dashed line). Because this
is the flux at the horizon, the flux of positively charged particles (red) is
always less than or equal to zero while the flux of negatively charged
particles is always greater than or equal to zero. This is because the flux is
proportional to $q u^r$, and $u^r<0$ for a particle falling into the horizon.

In the right columns of Figures \ref{Fig:FvQ} and \ref{Fig:Fvth} we show the
final cylindrical radius ($r_f \sin\theta_f$) of the positively charged and
negatively charged test particles with the same color scheme as in the left
column. The dotted black lines shows the initial radial distribution, chosen
to be at a constant spherical radius. We stop numerical integration when a
test charge either reaches $r=250M$, at which point we plot the spherical
radius to show that the charge has been expelled, or when the particle passes
within $0.01M$ of the horizon, or after a maximum time chosen to be
approximately the time for the particle to orbit the BH.

Figure \ref{Fig:FvQ} displays the horizon flux and final radii vs. the black
hole charge in units of the Wald charge. Figure \ref{Fig:Fvth} displays these
quantities vs. the initial starting position $\theta_i$.

The Wald argument on the poles can be readily seen from the top row of Figure
\ref{Fig:FvQ}. In the left panel we see that for $Q<Q_W$ the net flux into the
horizon at the pole is negative, meaning the BH is charging up and that this
flux is due entirely to positive charges. Above the Wald charge, the flux is
the opposite sign and due entirely to negative charges. The right panel in the
top row of Figure \ref{Fig:FvQ} shows that the this is caused by positive
charges falling in below the Wald charge and negative charges falling in above
the Wald charge. The middle row of Figure \ref{Fig:Fvth} shows clearly that at
the poles, when the BH is charged to the Wald charge, both signs of charge
fall in, resulting in zero charge accretion. This is in agreement with the
observation from Figure \ref{Fig:EdotB} that $\mathbf{E} \cdot \mathbf{B}$
changes sign at the poles at the Wald charge.

The bottom two rows in Figure \ref{Fig:FvQ}, as well as the top and bottom rows
of Figure \ref{Fig:Fvth} demonstrate that no charge is accreted away from the
poles, but for different reasons. At the equator, particles orbit stably. In
between the poles and the equator, particles are expelled or orbit stably
depending on $\theta_i$ and the value of $Q$, as discussed above.

It is interesting to note that at and above the Wald charge, there exists a
charged magnetosphere surrounding the BH in the equatorial regions.
Below the Wald charge, a region of opposite charges orbit at higher latitudes
than the equatorial charged region. The stability and long term existence of
these charged regions, however, is only determined here in the non-interacting
test-charge regime. The back reaction of this charged region on the
electromagnetic fields of the Wald solution must be included to discuss the
astrophysical importance of the Wald magnetosphere. For example,
charges being accelerated along B-field lines as they orbit the BH
could be in high enough quantity to screen the accelerated electric fields, thus
leading to the generation of a force-free magnetosphere. We discuss this
possibility in section \ref{S:FF}.

\subsection{Prospects for electromagnetic radiation}

Consider the charging process. As discussed above, at the poles, charges of
one sign are accelerated into the BH while charges of the opposite sign are
expelled from the BH. In the non-polar regions charges can orbit stably as
they are accelerated on oscillating orbits in the direction aligned with the
magnetic field and BH spin, or they can be continuously expelled along
B-field lines. All of these accelerated charges will emit electromagnetic (EM)
radiation.

EM radiation from the charging/discharging process could come from a few
mechanisms:  Dipole radiation from acceleration of ingoing and outgoing
charges and synchrotron and curvature radiation of stable orbits and ingoing
and outgoing orbits. The most promising of these processes for generating
bright EM signals is synchro-curvature radiation of particles expelled and
beamed away from the BH, as these experience the highest accelerations.  

Because positively charged particles, initially at rest, stay on cylindrical
orbits, there is a cylinder of initial coordinates given by $r_i
\sin{\theta_i} \leq r_+$ that will accelerate into the BH. There will also be
particles continuously accelerated away from the BH even at the Wald charge.
This can be seen in the right-middle panel of Figure \ref{Fig:Fvth}. The red
line shows that away from the poles positive particles are continuously
ejected from the system. While these particles will not be accelerated at the
poles at the Wald charge, they can be accelerated near to the BH and could
also contribute to a continuous signal of synchro-curvature radiation, if this
is a stable configuration. Here we estimate \textit{a maximum} power in EM
radiation that could be emitted by these charges during the
charging/discharging process of a Kerr BH, or even while the BH is charged at
the Wald charge.

The power generated by curvature radiation is
\begin{equation}
P_c = \frac{2}{3} q^2 c \frac{\gamma^4_q}{R^2_c}.
\end{equation}
As a charge is accelerated, radiation-reaction forces limit the maximum velocity $\mathbf{u}$ of the particle to where curvature radiation losses balance the power input from the electric field, $q \mathbf{E}\cdot\mathbf{u} = - P_c(\mathbf{u})$. 
For $|\mathbf{u}|\rightarrow c$ we approximate the radiation reaction condition as \citep{DL:2013}
\begin{equation}
E_Z = \frac{2}{3} q \frac{\gamma^4_q}{R^2_c}.
\end{equation}
We use $E_Z$ of the $Q_W=0$ Wald solution with a magnetic field due to a magnetic dipole
at a distance $R_c=20M$ and pole strength of $B_{NS}=10^{12}$G. Then at $r \sim
5M$ above the pole, the radiation reaction limited Lorentz factor of the charge is 
\begin{equation}
\gamma_q = 6.7 \times 10^7 \left(\frac{E_Z}{2.8 \times 10^9 \rm{statV} \rm{cm}^{-1}}\right)^{1/4} \left(\frac{R_c}{20M}\right)^{1/2},
\end{equation}
which represents a maximum Lorentz factor in the magnetosphere that is only weakly dependent on $E_Z$.

This large value of $\gamma$ is in agreement with previous studies
that estimate maximum particle Lorentz factors in magnetospheres sourced by NS
strength magnetic fields \citep{RudSuth:1975, DL:2016, DL:2013}. It is
physically justified by the large accelerating electric field. If instead we were to
solve for the velocity of an electron uniformly accelerated with
acceleration $a_e=(q/m) E_{Z}$, then we find that the velocity of the electron
in units of $c$ is $\beta_e = (a_e t/c)/\sqrt{1+(a_et/c)^2}$. This results in
the acceleration of the electron to $\gamma=10^7$ in approximately $10^{-10}$
seconds. Meaning that our approximate radiation-reaction velocities are
reached after the electron moves by of order a centimeter, a very short
distance compared to the scale of the system. For example, this is
$\approx10^{-5}(M/\Msun) \times$ smaller then the gravitational radius.

The maximum $\gamma_q$ can tell us the maximum power radiated by one charge via curvature radiation. The total number of charges is given by the Wald charge divided by the elementary charge,
\begin{equation}
\frac{Q}{e} = 4.54 \times 10^{33} \left(\frac{B_{NS}}{10^{12} G}\right)\left(\frac{R_{NS}}{r}\right)^3 \left(\frac{M}{10 \Msun}\right)^2.
\end{equation}
For charging and discharging this is the obvious choice. For a continual flux at the Wald charge, we choose this as a characteristic value because the repelling charge on the BH is likely comparable to the orbiting charge of opposite sign and the expelled charge. This at least describes a possible stable situation where the system remains electrically neutral at large distance.
Then the total power from curvature radiation during charge or discharge, or while the BH is charged at the stable Wald charge is,
\begin{equation}
P_c = 7.1 \times 10^{42} \rm{erg} \rm{s}^{-1} \left(\frac{B_{NS}}{10^{12} G}\right)^2 \left(\frac{R_c}{20M}\right)^{-6}  \left(\frac{M}{10 \Msun}\right)^2.
\label{Eq:ChargPow}
\end{equation}

Curvature radiation of this energy will spark a pair cascade
filling accelerating regions with an electron-positron pair plasma that will
eventually screen the accelerating fields \citep[see][and references
therein]{DL:2016}. This may not be an issue if we are only considering
charging and discharging of the BH because (dis)charging of the BH should take
of order the same time as the generation of the pair cascade (a light crossing
time of the system). For a continuous signal due to ejection of charges at the
Wald charge, however, the transition to a force-free magnetosphere may occur before the
near-merger separations needed for an observable signal, and hence the stable
fluxing scenario may be altered. 

This latter case, however, is similar to the 
that of a Pulsar magnetosphere, where particle production screens accelerating electric fields everywhere except for gaps where the force-free equations break down. In this case, the accelerating vacuum electric field can be reduced, and the region of acceleration is diminished to the size of the accelerating gap. While we plan to study this effect for the BHNS system with force-free and particle-in-cell simulations, for now we note that because our maximum Lorentz factor is only weakly dependent on the accelerating electric field, and because the gap height is likely larger than the acceleration distance estimated above, \citep[estimated to be of order the gravitational radius in Ref.][]{Chen+2018}, we expect these approximate results to be on the right track.

In the case of charging and discharging, the important question is when will
such a large magnetic field suddenly appear or disappear. For the case of the
inspiral of a BH and highly charged NS, the orbital decay timescale should
occur more slowly than the charging time of the BH and hence the charge of the
BH will increase at the rate that the magnetic field immersing the BH
increases. For a dipole magnetic field at time dependent distance $a(t)$ from
the BH, $Q_{\rm{Wald}}(t) \propto a^3(t) \propto t^{3/4}$ (assuming GW decay
of the binary \citep{Peters64}). At a critical separation the $\mathbf{E}$ and
$\mathbf{B}$ fields will become large enough for the pair cascade to spark.
Hence no sudden immersing of the BH in the electro-vacuum field of the NS is
expected.

The discharging case may only happen if the NS is swallowed. In this case the
destruction of the immersing field will also occur at either the light
crossing time of the BH \citep{BaumShap:2003}, or the resistive time of the
force-free magnetosphere that has been generated by pair production
\citep{LyutikovMckinney:2011}. In the former case, powerful radiation from
cleaning of the fields is generated, but at wavelengths of order the horizon
size. Such km wavelength radiation is not detectable as it has a frequency
below the plasma frequency of the galaxy \citep[see][]{DL:2016}. In the latter
case, a possible EM signature of a long-lived BH magnetosphere is discussed
in Refs. \cite{DL:2016} and \cite{LyutikovMckinney:2011}. If the Wald solution
is valid before the NS is swallowed, then a discharging signature similar to
the one discussed here may also accompany the merger.

In summary, a powerful EM signal with luminosity given by Eq.
(\ref{Eq:ChargPow}) could be generated by rapid charging or discharging of a
BH to or from the Wald charge, at the BH poles. A similar luminosity could
also be generated by a BH at the stable Wald charge from the continual
expulsion of charges away from the poles. In a related scenario, a
magnetosphere sourced by the spinning, charged BH could result in emission
mechanisms similar to that of a pulsar. A more refined
prediction for detection would benefit from an understanding of the back
reaction of charge acceleration in the Wald field.

\section{Charged Force-Free Solutions}
\label{S:FF}

Force-free solutions are notoriously hard to come by, and we reserve the attempt at a force-free set-up for another work (see e.g.\ \cite{Komissarov:2002my,Gralla:2014yja} for formal aspects of force-free electrodynamics, and \cite{Palenzuela:2010xn,Contopoulos:2012py,Brennan:2013kea,Lupsasca:2014pfa,Zhang:2014pla,Compere:2015pja,Gralla:2016jfc,Li:2017qzu} for studies of force-free fields in BH spacetimes). However, the reader might be concerned, as we were, that force-free solutions somehow ensure an uncharged BH. Although this would not prohibit the charge up during the vacuum phase, it would be worth knowing if charge could be sustained. We consider the Blandford-Znajek (BZ) split monopole on a BH to show that the BH retains charge. A related analysis for non-rotating BHs was performed in \cite{Lee:2000tm} based on a force-free solution derived in \cite{Ghosh:1999in}.

\subsection{Charge of the Blandford--Znajek split monopole: Gauss' law}

Although our main interest is in computing the electric charge enclosed within the horizon of the BH, it is instructive to do something slightly more general and calculate the charge inside an arbitrary sphere of radius $R$, defined as the 2-surface $r=R$ in Boyer--Lindquist coordinates. Applying Gauss' law the charge $Q$ is given by
\begin{equation}
4\pi Q(R)=\int_{r=R}\star F=\int_{r=R}(\star F)_{\theta\phi}\,d\theta\wedge d\phi\,.
\label{Eq:GL}
\end{equation}
The Maxwell equations and force-free (FF) conditions are
\begin{eqnarray}
D\cdot F &=&  \J\,, \\
F\cdot \J &=& 0\,,
\end{eqnarray}
where the second equation clearly matches the case of a test particle, for which  $\J=q u$, with zero Lorentz force in Eq.\ (\ref{Eq:LF}). For an axisymmetric, stationary current, a function $\omega(r,\theta)$ can be defined through the FF conditions \cite{Blandford:1977ds}
\begin{equation}
A_{t,r}=-\omega A_{\phi,r}\,,\qquad A_{t,\theta}=-\omega A_{\phi,\theta}\,.
\end{equation}
The Hodge dual $(\star F)_{\alpha\beta}=(1/2)\varepsilon_{\alpha\beta\mu\nu}F^{\mu\nu}\,$ (with 
$\varepsilon_{tr\theta\phi}=-\sqrt{-g}$) gives
\begin{equation} \label{Eq:starF}
\begin{split}
(\star F)_{\theta\phi}&=-\frac{1}{2}\,\varepsilon_{\theta \phi \mu\nu}F^{\mu\nu}=-\sqrt{-g}\,F^{tr}\\
&=-\sqrt{-g}\,g^{rr}A_{\phi,r}(\omega g^{tt}-g^{t\phi})\,.
\end{split}
\end{equation}

The BZ split monopole solution corresponds to
\begin{equation} \label{eq:BZ_EMfield}
\begin{split}
\omega&=\frac{a}{8M^2}\left(1+O\left(\frac{a}{M}\right)^2\right)\,,\\
A_{\phi}&=-C|\cos\theta|+\frac{Ca^2}{M^2}\,f(r)\sin^2\theta|\cos\theta|+O\left(\frac{a}{M}\right)^4\,,
\end{split}
\end{equation}
where $C$ is just a constant gauging the strength of the split monopole and $f(r)$ is the dimensionless function \cite{Pan:2015haa}
\begin{equation}
\begin{aligned}
f(r)&=\frac{1+3(r/M)-6(r/M)^2}{12}\,\ln\left(\frac{r}{2M}\right)+\nonumber \\
&\frac{11}{72}+\frac{M}{3r}+\frac{r}{2M}
-\frac{r^2}{2M^2}+
\frac{r^2(2r-3M)}{8M^3}\times \nonumber \\
&\left[{\mathrm{Li}}_2\left(\frac{2M}{r}\right)-\ln\left(1-\frac{2M}{r}\right)\ln\left(\frac{r}{2M}\right)\right]\,,
\end{aligned}
\end{equation}
and
\begin{equation}
{\mathrm{Li}}_2(x)=-\int_0^1\frac{\ln(1-tx)}{t}\,dt\,.
\end{equation}

Notice that the absolute value in $A_{\phi}$ is enforced so that the radial magnetic field is odd upon reflection about the equator,
\begin{equation}
\begin{split}
B^r&=-\frac{1}{2}\varepsilon^{r \nu  \alpha \beta} F_{\alpha \beta} u_\nu  \propto A_{\phi, \theta}\left (  u_t+\omega u_\phi \right )  \\
\end{split}
\end{equation}
and $A_{\phi, \theta}$ clearly changes sign under $\theta\to\pi-\theta$. In other words, it's a split monopole. Another check one can make is to compute the magnetic charge on the BH (using $F$ instead of $\star F$ in Gauss' law) and verify that it's zero by symmetry. If we ask an observer at rest very far from the BH, they see a split monopole field that goes like $B^r\sim \pm C/r^2$, so $C$ has the meaning of a magnetic charge.

\begin{figure}
\centering
\includegraphics[width=\linewidth]{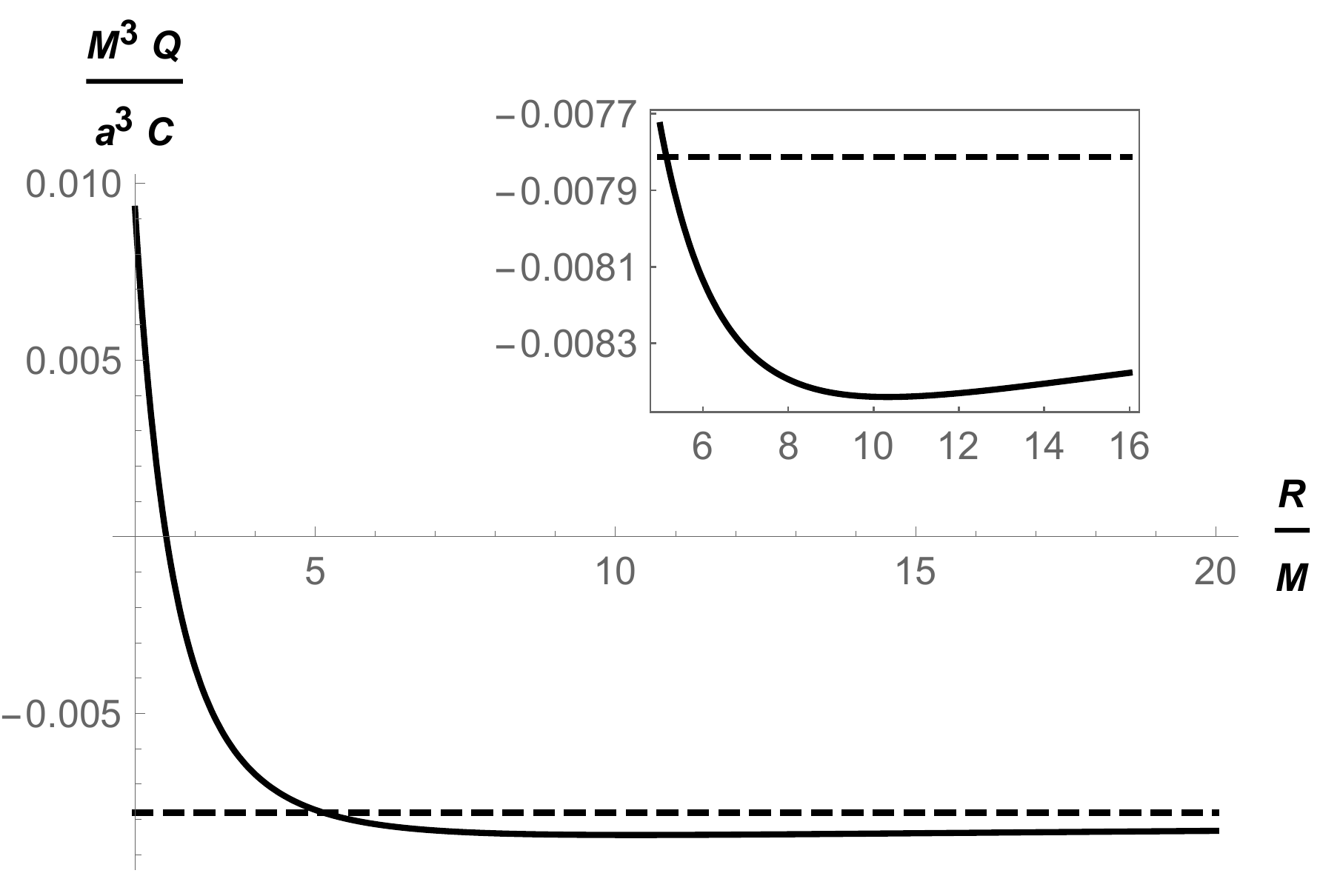} 
\caption{Electric charge $Q$ enclosed within a sphere of radius $R$, plotted as a function of $R/M$ and normalized by $Ca^3/M^3$. The dashed horizontal line corresponds to the value $-1/128$ that the curve approaches to at infinity. The inset is a detail of the same plot showing the minimum of $Q$ at $R\simeq10 M$.}
\label{fig:BZ_Qplot}
\end{figure}

Substituting \eqref{eq:BZ_EMfield} in \eqref{Eq:starF} we have
\begin{equation}
\begin{split}
(\star F)_{\theta\phi}&=\frac{Ca^3}{M^4}\,r^2f'(r)\sin^3\theta|\cos\theta|\left(\frac{1}{8}-\frac{2M^3}{r^3}\right)\\
&\quad\times\left(1+O\left(\frac{a}{M}\right)^2\right)\,.
\end{split}
\end{equation}
Finally, the angular integral yields
\begin{equation}
\int \sin^3\theta|\cos\theta|\,d\theta d\phi=\pi\,,
\end{equation}
and we arrive at
\begin{equation} \label{eq:BZ_fullQ}
Q(R)=\frac{Ca^3}{4M^4}\,R^2f'(R)\left(\frac{1}{8}-\frac{2M^3}{R^3}\right)\left(1+O\left(\frac{a}{M}\right)^2\right)\,.
\end{equation}
Two interesting values of $R$ are the horizon $r_{+}\simeq2M$ and infinity. We find
\begin{equation}
\begin{split}
Q(r_{+})&=\frac{Ca^3}{8M^3}\left(\frac{61}{24}-\frac{\pi^2}{4}\right)\left(1+O\left(\frac{a}{M}\right)^2\right)\,,\\
Q(\infty)&=-\frac{Ca^3}{128M^3}\left(1+O\left(\frac{a}{M}\right)^2\right)\,.\\
\end{split}
\end{equation}
The standard application of Gauss's law in an asymptotically flat spacetime is from far away.
Interestingly, the charge is not the same at infinity as at the horizon, suggesting the magnetosphere is charged as well. This suggests that the magnetosphere carries positive charge $Q_M\sim -(5/4)Q_{r_+}$. Fig.\ \ref{fig:BZ_Qplot} shows a plot of the charge as a function of the Gaussian surface radius $R$ at the order in $a/M$ to which we are working.

\subsection{Which observer sees charge $Q$?}

Notice that Eq.\ \eqref{Eq:GL}, can be related to the naive form of Gauss's Law:
\begin{equation}
4\pi Q=\int_{r=r_{+}}\star F=\int_{r=r_{+}}\left (E\cdot n\right ) \sqrt{g_{2D}}\ d\theta d\phi\,,
\label{Eq:Step0}
\end{equation}
but does not necessarily correspond to any fields measured by timelike observers on the horizon.
In other words, with 
\begin{eqnarray}
E\cdot n = (F\cdot u)\cdot n \label{Eq:Step1}\,,
\end{eqnarray}
we can use the results of the previous section to glean the $u$ required to measure the field for a normal to a sphere
\begin{equation}
n^\mu=g_{rr}^{-1/2}(0,1,0,0)
\end{equation}
and then we equate the integrand on the RHS of Eq.\ \eqref{Eq:Step0} using Eq.\ \eqref{Eq:Step1} to find
\begin{equation}
\begin{split}
E_r g_{rr}^{-1/2} = g_{rr}^{-1/2} F_{r\mu} u^\mu &= N F^{tr}\\
&=-Ng^{rr}\left (F_{r\mu} g^{\mu t}\right )\,,
\end{split}
\end{equation}
where we used Eq.\ \eqref{Eq:starF} and 
\begin{equation}
N=\frac{\sqrt{-g}}{\sqrt{g_{2D}}}=\frac{\Sigma}{((r^2+a^2)^2-\Delta a^2\sin^2\theta)^{1/2}}\,.
\end{equation}
At the horizon,
\begin{equation}
N_+= \frac{\Sigma_+}{2Mr_+}\, .
\end{equation}
There is an observer that satisfies the above with 4-velocity of the form
\begin{equation}
\begin{split}
u^\mu 
& = -N g_{rr}^{-1/2} g^{\mu t} \\ & 
= N g_{rr}^{-1/2}\left ( - g^{tt},0 , 0, - g^{t\phi}\right ) \, .
\end{split}
\end{equation}

Mercifully, our observer has timelike norm at the horizon. Exploiting relations Eq.\ (\ref{Eq:U3})
\begin{eqnarray}
u \cdot u &=& N^2 \frac{\Delta}{\Sigma}  \left (g^{tt} g^{tt} g_{tt}+2g^{tt}g^{t\phi}g_{t\phi}
 +g^{t\phi}g^{t\phi}g_{\phi\phi} \right) \nonumber \\
 &=& N^2\frac{\Delta}{\Sigma}  \left (g^{tt} \left (g^{tt} g_{tt}+g^{t\phi}g_{t\phi}\right )
 +g^{t\phi}\left (g^{tt}g_{t\phi}+g^{t\phi}g_{\phi\phi} \right)\right ) \nonumber \\
 &=& N^2\frac{\Delta}{\Sigma}  g^{tt}  =-1\,,
 \end{eqnarray}
as desired.

Notice that we can neatly verify that $E\cdot B=0$.  To do so we note that for this $u$ the non-zero $E$-field components are $E=(0,E_r, E_\theta, 0 )$ and so 
 \begin{equation}
 E\cdot B= E_r B^r +E_\theta B^\theta =0
 \label{Eq:EB}
 \end{equation}
 Now for this $A$ we can also express $E_r$ as
 \begin{equation}
 \begin{split}
 E_r &= F_{r\phi}\left (-\omega u^t+u^\phi \right ) \\
 & = A_{\phi,r}\left (\omega u^t-u^\phi \right ) 
 \end{split}
 \end{equation}
Using that $u_\phi =g_{\phi \mu} u^\mu =0$ according to Eqs.\ (\ref{Eq:U3}), we compare $E_r$ to 
 \begin{equation}
 \begin{split}
\sqrt{-g} B^\theta &= -\frac{1}{2}\epsilon^{\theta \nu  \alpha \beta} F_{\alpha \beta} u_\nu \\
& = \frac{1}{2}\epsilon^{\theta i j} F_{ij} u_t \\
& = - A_{\phi, r} u_t \, ,
 \end{split}
 \end{equation}
 to find
 \begin{equation}
 E_r= - \sqrt{-g}B^\theta \left (\frac{\omega u^t-u^\phi  }{u_t}\right ) \, .
 \end{equation}
 Similarly
 \begin{equation}
 E_\theta =  \sqrt{-g}B^r \left (\frac{\omega u^t-u^\phi  }{u_t}\right ) \, .
 \end{equation}
 Putting these in Eq.\ (\ref{Eq:EB}) immediately yields 0.

\subsection{Neutron-star pulsar charge}

Although our goal was to provide evidence that a BH surrounded by a force-free magnetosphere can support charge, it is instructive to compare this situation with that of a NS. As a crude model of a NS pulsar we consider a magnetic dipole in the Goldreich--Julian set-up \cite{Goldreich:1969sb}, i.e.\ with a co-rotating magnetosphere within the light cylinder and ignoring gravitational effects (see \cite{1975ApJ...201..783C} for a rigorous analysis of the same model; see also \cite{Gralla:2016fix} for a study of more general models of pulsar magnetospheres).

The vanishing of the Lorentz force for a charge with 3-velocity $\vec{v}$ relates the electric and magnetic fields as
\begin{equation}
\vec{E}=-\vec{v}\times\vec{B}\,.
\end{equation}
In a co-rotating magnetosphere the charge's velocity is given by $\vec{v}=\Omega_{NS}r\sin\theta\,\hat{\phi}$, with $\Omega_{NS}$ the star's angular velocity, and as we mentioned the magnetic field is idealized as that of a magnetic dipole with moment $\vec{m}\equiv m\hat{z}$. Then
\begin{equation}
\vec{E}=\frac{\Omega_{NS}m}{r^2}\,\sin\theta\left(\sin\theta\,\hat{r}-2\cos\theta\,\hat{\theta}\right)\,.
\end{equation}
Using Gauss' law we obtain the following result for the charge contained in the NS:
\begin{equation}
\begin{split}
4\pi Q_{NS}&=\int_{r=R_{NS}}\vec{E}\cdot\hat{r}\,r^2\sin\theta\,d\theta d\phi\\
&=\frac{8\pi}{3}\,\Omega_{NS}m\,,
\end{split}
\end{equation}
which is in fact independent of the radius of the Gaussian sphere. Thus, in contrast to the BH case we focused on, the magnetosphere surrounding the star carries zero net charge in this simplified model.

We can define the characteristic magnetic field strength, $B_{NS}$, of the NS via the relation $m\equiv B_{NS}R_{NS}^3$, so that
\begin{equation}
Q_{NS}=\frac{2}{3}\,\Omega_{NS}B_{NS}R_{NS}^3\,.
\end{equation}
The numerical factor is of course rather meaningless given the simplifications we have made, but we may expect this result to give a correct order of magnitude.

Comparing with an estimate of the peak Wald charge expected before the merger with a maximally spinning BH, $Q_W\sim B_{NS}M^2$, we find,
\begin{equation}
\frac{Q_{NS}}{Q_W}\sim10^{-4}\left(\frac{\Omega_{NS}}{1\,{\rm s^{-1}}}\right)\left(\frac{R_{NS}}{10^6\,{\rm cm}}\right)^3\left(\frac{10\,M_{\odot}}{M_{BH}}\right)^2\,.
\end{equation}
This shows that the increase in the charge of the BH upon swallowing the NS is likely to be negligible compared to the maximum charge accreted during the inspiral phase as quantified by the Wald charge.

\section{Summary}
\label{S:Summary}

The wealth of information gained from the NS/NS merger GW170817 and GRB170817
speaks compellingly to the prodigious importance of electromagnetic
counterparts to gravitational-wave signals. Arguing against convention, in
this paper we have put forth the idea that a valuable counterpart to a BH/NS
merger may exist by leveraging the charge BHs can support.

BH charge is typically dismissed in astrophysical settings based on the
expectation that charge will be both negligibly tiny and/or extremely short-
lived. The presumption that charge is short-lived is countered by the Wald mechanism
--- a rotating BH embedded in an external magnetic field will accrete a stable
net charge. Further, the charge need not be tiny given the magnitude of strong
NS B-fields and rather could be relevant to observations.

A simple estimate of the magnetic field created by the BH as charge reaches
its maximum value immediately before the merger with a strongly magnetized NS
gives $B_{BH}\sim (a/M)^2 B_{NS}/2$, comparable to the NS magnetic field for
highly spinning BHs. As found observationally, and through theoretical
investigation, whether or not a NS can generate a magnetosphere and produce
pulsar emission depends on the spin period of the NS. For example,
\citet{Sturrock:1971} and \citet{RudSuth:1975} calculate that a NS must have a
period shorter than $\sim1.7 (B_{NS}/10^{12})^{8/13}$ seconds to sustain
charge acceleration across a vacuum gap and hence the pulsar magnetosphere.
The spin period of a maximally spinning, $10\Msun$ BH is of order
milliseconds. Hence, if the analogy can be applied to the BH-pulsar case, this
means that BHs sourcing magnetic fields above $10^7-10^8$~G should be able to
sustain a magnetosphere, and possibly drive an emission mechanism similar to
that of the pulsar case. Promisingly, recent numerical work has employed
particle-in-cell simulations of BH magnetospheres finding that small polar
gaps, analogous to the NS-pulsar case, can be opened and result in particle
acceleration \citep[see][and references therein]{Chen+2018}. For mergers involving NS surface magnetic fields of
$B_{NS}\sim10^{12}$~G, the final $\sim 20M$ of inspiral, would allow the BH to
source a magnetic dipole field of $\gtrsim 10^{8}$~G, above the pulsar limit. 

It should be emphasized, however, that this ``black hole pulsar,'' as we have
called it, has an essential difference relative to a NS: its magnetic field is
created by a rotating electric charge, unlike the star's intrinsic dipole
field. After all, a co-rotating observer sees only an electric field due to
the charge on the BH. Granted, the pulsar features of such a BH may be hard to
observe given its short lifetime and their scarcity within galactic distances.
And any detailed predictions would require an analysis of the generalization
to a time-dependent, non-uniform external magnetic field.

Another source of luminosity can stem from the acceleration of charges
surrounding a BH, which we have shown is not precluded by the stability of the
net charge. The vacuum situation we considered suggests some interesting
properties, as demonstrated in the complex, likely chaotic, dynamics. And even
though our estimates for the emitted power via curvature radiation are large
enough to be interesting (of order kilonova luminosities
\citep[\textit{e.g.},][]{MetzgerKilonova+2010}), a more accurate prediction would pose similar
difficulties that make the NS-pulsar studies so challenging.

We began with vacuum solutions, however the BH may well create its own force-free magnetosphere. If that transpires, we can ask whether a BH charge and its associated effects should then be dismissed. Again, against expectation, we showed that a BH enclosed by a force-free magnetosphere does in fact carry charge. Still, the situation we focused on --- the Blandford--Znajek split monopole --- is an approximate force-free solution valid for small $a/M$, leading to a correspondingly small electric charge. We believe nonetheless that this outcome is interesting enough to motivate a more thorough numerical study on the existence of electric charge in force-free BH magnetospheres. We note that, without speculating on the origin of charge on BH/BH pairs, the same mechanisms would be at work to illuminate these systems if they exist.

Finally, it is interesting to speculate, should an electromagnetic counterpart
to a BH/NS or BH/BH merger be observed, about the prospects of testing fundamental
physics. The no-hair theorem immediately comes to mind, as the detection of a
BH pulsar could in principle be sensitive to an intrinsic magnetic (dipole or
higher) moment. A positive detection could also be used to constrain modified
gravity theories in which the analogue of the Wald mechanism
\citep[\textit{e.g.},][]{PCWald:2017} might differ significantly from that in
general relativity.

\acknowledgements

The authors would like to thank Samuel Gralla, Bob Penna, Rhondale Tso, and Bob Wald for
useful input and discussions. JL thanks Science Sandbox of the Simons
Foundation for generous support of the Science Studios at Pioneer Works.
Financial support was provided by NASA through Einstein Postdoctoral
Fellowship award number PF6-170151 (DJD), and by the European Research Council
under the European Community's Seventh Framework Programme, FP7/2007-2013
Grant Agreement no.\ 307934, NIRG project (SGS).

\appendix
\section{}
\label{App:A}

Calculations are greatly facilitated by a list of clean relationships among metric quantities. We compile those relations here.

Again, in Boyer-Lindquist coordinates:
\begin{eqnarray}
ds^2 = &-& \left (1-\frac{2Mr}{\Sigma}\right ) dt^2 +\frac{\Sigma}{\Delta} dr^2 + \Sigma d\theta^2 \nonumber \\
&+& \frac{(r^2+a^2)^2-\Delta a^2\sin^2\theta}{\Sigma}\sin^2\theta d\phi^2 \nonumber \\
&-& \frac{4Mar\sin^2\theta}{\Sigma} dt d\phi\,,
\label{Eq:BL}
\end{eqnarray}
with
\begin{eqnarray}
\Sigma & =& r^2 + a^2\cos^2\theta \,, \nonumber \\
\Delta &=& r^2+a^2 - 2Mr \,.
\end{eqnarray}

Useful metric quantities are
\begin{equation}
\begin{split}
\sqrt{-g}&=\Sigma\,\sin\theta\,,
\\
g^{rr}&=\frac{\Delta}{\Sigma}\,,
\\
g^{\theta\theta}&=\frac{1}{\Sigma}\,,
\\
g^{tt}&=-\left (\frac{(r^2+a^2)^2-\Delta a^2\sin^2\theta}{\Delta\Sigma}\right )\,,\\
g^{t\phi}&=-\frac{2Mar}{\Delta\Sigma}\,, \\
g^{\phi \phi} &= \frac{\Delta -a^2\sin^2\theta}{\Sigma\Delta \sin^2\theta}\,.
\end{split}
\end{equation}

Other useful equalities:
\begin{equation}
\begin{split}
\Sigma &= \Delta +2Mr-a^2\sin^2\theta  \,,\\
g_{tt}g_{\phi\phi}-g_{t\phi}^2 &= -\Delta \sin^2\theta \,,\\
g^{tt} &= \frac{g_{\phi\phi}}{(g_{tt}g_{\phi\phi}-g_{t\phi}^2)} \,,\\
g^{t\phi} &= \frac{-g_{t\phi}}{(g_{tt}g_{\phi\phi}-g_{t\phi}^2)}  \,,\\
g^{\phi\phi} &= \frac{g_{tt}}{(g_{tt}g_{\phi\phi}-g_{t\phi}^2)}  \,.
\label{Eq:Use}
\end{split}
\end{equation}
Also, by the definition of an inverse
\begin{eqnarray}
g_{tt}g^{tt}+g_{t\phi}g^{t\phi} & =& 1 \,,\nonumber \\
g_{tt}g^{t\phi} + g_{t\phi}g^{\phi\phi} &=& 0 \,,\nonumber \\
g^{tt}g_{t\phi} + g^{t\phi}g_{\phi\phi} &=& 0\,.
\label{Eq:U3}
\end{eqnarray}


\bibliographystyle{apsrev4-1}
\bibliography{BlackHoleCharge_biblio}

\end{document}